\documentclass[%
superscriptaddress,
preprint,
preprintnumbers,
 amsmath,amssymb,
 aps,
]{revtex4-2}
\usepackage[colorlinks=true, allcolors=blue]{hyperref}
\usepackage{graphicx}
\usepackage{dcolumn}
\usepackage{bm}
\usepackage{xcolor} 
\usepackage{natbib}
\usepackage{physics}

\usepackage{gensymb}

\begin{document}


\title{Field-induced first order transitions and phase coexistence in the Kitaev quantum spin liquid candidate BaCo$_{2}$(AsO$_{4}$)$_{2}$}

\author{Tanner J. Legvold}
\thanks{These authors contributed equally.}
\thanks{deceased}
\affiliation{Department of Physics and Astronomy, Rice University, Houston, TX USA 77005} 
\author{Bin Gao}
\thanks{These authors contributed equally.}
\affiliation{Department of Physics and Astronomy, Rice University, Houston, TX USA 77005} 
\affiliation{Rice Laboratory for Emergent Magnetic Materials and Smalley-Curl Institute, Rice University, Houston, TX 77005, USA}
\author{Rong-Zhu Lin}
\thanks{These authors contributed equally.}
\affiliation{Department of Physics and Astronomy, Rice University, Houston, TX USA 77005} 
\affiliation{Rice Laboratory for Emergent Magnetic Materials and Smalley-Curl Institute, Rice University, Houston, TX 77005, USA}
\affiliation{Department of Physics and Center for Quantum Frontiers of Research \& Technology (QFort), National Cheng Kung University, Tainan,Taiwan}
\author{Violet Williams}
\affiliation{Department of Physics, Boston College, Boston, MA USA 02467}
\author{Tong Chen}
\affiliation{Department of Physics and Astronomy, Rice University, Houston, TX USA 77005} 
\affiliation{Rice Laboratory for Emergent Magnetic Materials and Smalley-Curl Institute, Rice University, Houston, TX 77005, USA}
\author{Dehong Yu}
\affiliation{Australian Nuclear Science and Technology Organization (ANSTO), Lucas Heights, New South Wales, Australia}
\author{Chien-Lung Huang}
\affiliation{Department of Physics and Center for Quantum Frontiers of Research \& Technology (QFort), National Cheng Kung University, Tainan,Taiwan}
\author{Gage Eichman}
\affiliation{Department of Physics and Astronomy, Rice University, Houston, TX USA 77005} 
\author{Renjie Luo}
\affiliation{Department of Physics and Astronomy, Rice University, Houston, TX USA 77005} 
\author{Benedetta Flebus}
\affiliation{Department of Physics, Boston College, Boston, MA USA 02467}
\author{Pengcheng Dai}
\email{pdai@rice.edu}
\affiliation{Department of Physics and Astronomy, Rice University, Houston, TX USA 77005} 
\affiliation{Rice Laboratory for Emergent Magnetic Materials and Smalley-Curl Institute, Rice University, Houston, TX 77005, USA}
\author{Douglas Natelson}
\email{natelson@rice.edu}
\affiliation{Department of Physics and Astronomy, Rice University, Houston, TX USA 77005}
\affiliation{Rice Center for Quantum Materials, Smalley-Curl Institute, Rice University, Houston, TX USA 77005}
\affiliation{Department of Electrical and Computer Engineering, Rice University, Houston, TX USA 77005}
\affiliation{Department of Materials Science and NanoEngineering, Rice University, Houston, TX USA 77005}

\date{\today}

\begin{abstract}
BaCo$_2$(AsO$_4$)$_2$ (BCAO) is an insulating Kitaev quantum spin liquid candidate with a rich low-temperature phase diagram. Below 5 K, it exhibits double-zigzag magnetic order. Upon application of an in-plane magnetic field, the magnetic structure first transforms into an up-up-down (UUD) state near 0.15 T and then enters a fully spin-polarized ferromagnetic (FM) state near 0.5 T. In the narrow field regime close to the polarized phase, a finite residual thermal conductivity has been reported, suggesting a possible field-induced quantum spin liquid phase. Using neutron scattering, we show that the field-induced double-zigzag-to-UUD transition near 0.15 T is accompanied by the emergence of a cluster of localized UUD spin excitations that coexist with conventional spin waves. Upon further increasing field, BCAO undergoes a first-order UUD-to-FM transition with coexistence of UUD and FM phases, accompanied by a sharp response in the spin Seebeck coefficient. These results do not support a quantum spin liquid scenario near the UUD-to-FM critical field. Instead, modeling indicates that the broad excitations arise from bound spin-flip pairs, while a low-lying dispersive branch near the FM phase boundary carries the same sign of magnetization as the FM order. These excitations naturally account for the observed sign of the spin Seebeck response and are likely relevant to the  thermal conductivity near the critical field.
\end{abstract}

\maketitle

\section{Introduction}

\begin{figure}[t]
   \includegraphics[width=\linewidth]{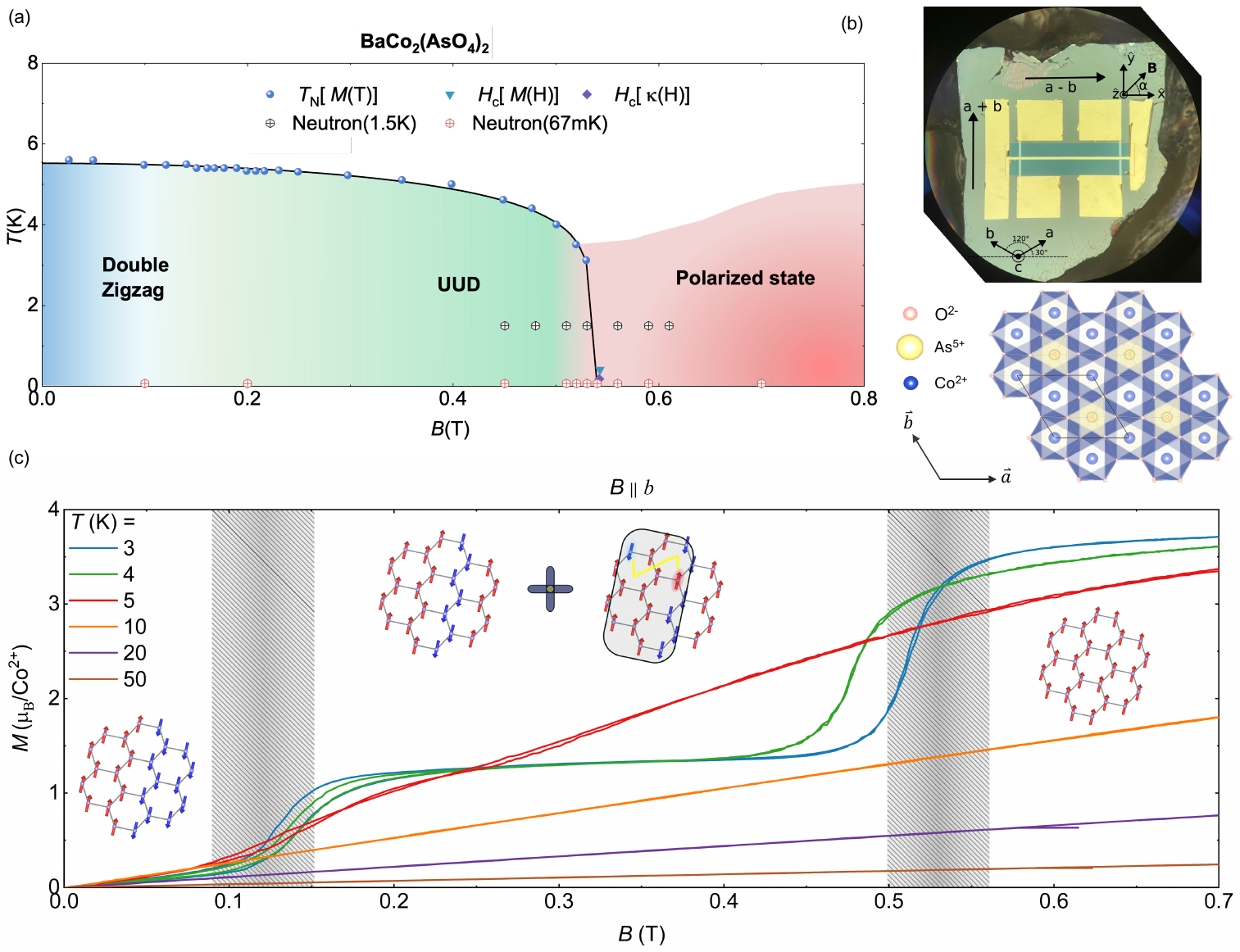}
    \centering
    \caption{\textbf{Crystal structure and magnetic phase diagram of $\text{BaCo}_2(\text{AsO}_4)_2$ (BCAO).} (a) Magnetic phase diagram of BCAO. (b) Photograph of a representative single crystal, alongside its reciprocal lattice and honeycomb crystal structure. (c) Isothermal magnetization as a function of magnetic field ($B \parallel b$). The grey hatched areas denote field-induced phase transitions. Insets illustrate the proposed spin configurations: the low-field double zigzag phase (left), the intermediate up-up-down (UUD) phase (center), and the high-field fully polarized state (right).}
    \label{fig:bcao_phase_diagram}
\end{figure}

The quantum spin liquid (QSL) state is an intriguing target of investigation, motivated by the possibility that strong exchange interactions and quantum fluctuations in a geometrically frustrated lattice can suppress long-range magnetic order and produce an emergent quantum entangled state with fractionalized excitations, such as neutral fermionic spinons \cite{broholm20qsl,zhou17qslrmp,savary16qsl}. Among QSL candidate materials, effective spin-$1/2$ honeycomb-lattice magnets are particularly attractive, including the exactly solvable Kitaev QSL model \cite{Takagi2019KitaevQSL,3m4m-3v59}. BaCo$_2$(AsO$_4$)$_2$ (BCAO), which contains a hexagonal lattice of Co$^{2+}$ ions, has attracted considerable recent attention \cite{doi:10.1126/sciadv.aay6953} because of its structural similarity to the well-studied Kitaev candidate $\alpha$-RuCl$_3$ \cite{Takagi2019KitaevQSL,3m4m-3v59}. BCAO exhibits a rich low-temperature phase diagram in modest magnetic fields, reflecting competition among several ordered states (Figs. 1a,1b) \cite{doi:10.1126/sciadv.aay6953}. At zero field, its magnetic structure can be described as quasi-ferromagnetic (FM) zigzag pseudo-chains running along the $b$ axis, with a phase angle between adjacent pseudo-chains, forming the double-zigzag structure shown in Fig.~1c \cite{10.1016/0378-4363(77)90635-0,REGNAULT1977660}. Upon applying a magnetic field along the $b$ axis, this double-zigzag order transforms into a field-induced up-up-down (UUD) state near 0.15 T, which is subsequently suppressed in favor of a field-polarized FM state near 0.5 T (Fig. 1c) \cite{doi:10.1126/sciadv.aay6953,10.1016/0378-4363(77)90635-0,REGNAULT1977660,REGNAULT1979194,REGNAULT2006425,REGNAULT2018e00507}.

Although the field-induced UUD phase is well established over the approximate field range $0.15 \leq B \leq 0.5$ T, previous elastic and inelastic neutron scattering (INS) experiments on triple-axis spectrometers did not determine the microscopic Hamiltonian or the processes driving the double-zigzag-to-UUD and UUD-to-FM transitions \cite{REGNAULT2018e00507}. In particular, no clear evidence was found for low-energy coherent spin waves emanating from the magnetic Bragg peaks associated with UUD order \cite{REGNAULT2006425}. The nature of the high-field transition has become especially important because low-temperature thermal conductivity measurements in the narrow field region where UUD order gives way to the field-polarized FM state revealed an apparent nonzero intercept of $\kappa/T$ as $T \rightarrow 0$ K \cite{tu25bcao}. In an insulator, a finite linear contribution to thermal conductivity in the zero-temperature limit is often taken as evidence for mobile fermionic quasiparticles with a Fermi surface, termed a spinon Fermi surface, analogous to the thermal transport of degenerate electrons in metals \cite{broholm20qsl,zhou17qslrmp,savary16qsl,li20thermalxport}.

The microscopic interpretation of BCAO remains unsettled. Terahertz spectroscopy has reported a field-induced excitation continuum consistent with a QSL-like state \cite{zhang23bcaothz}. By contrast, INS measurements in the fully field-polarized state found that the FM spin waves are better described by an XXZ-$J_1$-$J_3$ model than by a bond-dependent Kitaev-type $JK\Gamma\Gamma^\prime$ Hamiltonian \cite{doi:10.1073/pnas.2215509119}. More recent analysis of existing INS data \cite{REGNAULT2018e00507,doi:10.1073/pnas.2215509119}, however, argues for the presence of bond-dependent interactions, keeping open the possibility of proximity to Kitaev/QSL physics \cite{maksimov25kitaev}. Recent theoretical work has also shown that quantum fluctuations can stabilize double-zigzag order within an isotropic $J_1-J_3$ model \cite{ralko2026orderbydisorder}, further illustrating the ambiguity in identifying the minimal microscopic Hamiltonian for BCAO. These unresolved issues motivate a direct microscopic investigation of the field-induced excitations and phase evolution across the double-zigzag, UUD, and field-polarized regimes.

Spin transport techniques, including the spin Seebeck effect (SSE) \cite{kikkawa23sse} and spin Hall magnetoresistance (SMR) \cite{chen16smr}, provide sensitive access to spin-carrying excitations in magnetic insulators. Widely applied to magnetically ordered systems and paramagnets, these techniques have also been proposed as probes of fractionalized excitations in quantum magnets \cite{kikkawa23sse,kato25sse}. Combining SSE and SMR measurements with high-resolution INS on high-quality single crystals therefore offers a powerful approach for characterizing low-energy spin excitations and their ability to carry spin angular momentum in candidate QSLs.

Here we investigate high-quality BCAO single crystals using neutron scattering, SMR, and SSE measurements. Elastic neutron scattering reveals magnetic order throughout the low-temperature field phase diagram, while INS finds no evidence for a broad spinon-like continuum in the field regime where anomalous thermal transport has been reported. Instead, both the field-induced double-zigzag-to-UUD and UUD-to-FM transitions exhibit phase coexistence, indicating first-order transitions. The double-zigzag-to-UUD transition near 0.15 T is accompanied by the emergence of a localized excitation, which is subsequently suppressed at the UUD-to-FM transition near 0.5 T. SSE measurements reveal a sharp response only in the UUD/FM coexistence regime, with a sign indicating transport of magnetization aligned with the applied field. In this coexistence regime, localized bound spin-flip excitations can become mobile, and dispersive UUD spin excitations coupled to phonons can propagate while carrying field-aligned magnetization. These excitations must be considered when analyzing the previously reported anomalous low-temperature thermal transport \cite{tu25bcao}. Theoretical modeling consistent with the observed phase diagram and measured excitations requires strong Kitaev-like bond-dependent exchange interactions.

\section{Experimental Results}

Single crystals of BCAO were grown by the flux method \cite{doi:10.1126/sciadv.aay6953}, and their orientations were confirmed by Laue diffraction. INS experiments were performed on PELICAN, a time-of-flight cold-neutron spectrometer at the Open Pool Australian Lightwater (OPAL) reactor of the Australian Nuclear Science and Technology Organisation (ANSTO). The instrument was operated with a fixed incident energy of 3.71 meV, giving an energy resolution of approximately 0.15 meV at the elastic line. Multiple BCAO single crystals were co-aligned and mounted on an aluminum plate, then cooled in a dilution-refrigerator insert inside a 7-T superconducting magnet. The samples were aligned in the $[H,-H,0]\times[0,0,L]$ scattering plane, with the magnetic field applied in plane along the $[H,H,0]$ direction. The scattering intensity was placed on an absolute scale using a vanadium standard. Data reduction and visualization were performed using the Horace software package.

\begin{figure}[h!]
\includegraphics[width=15cm]{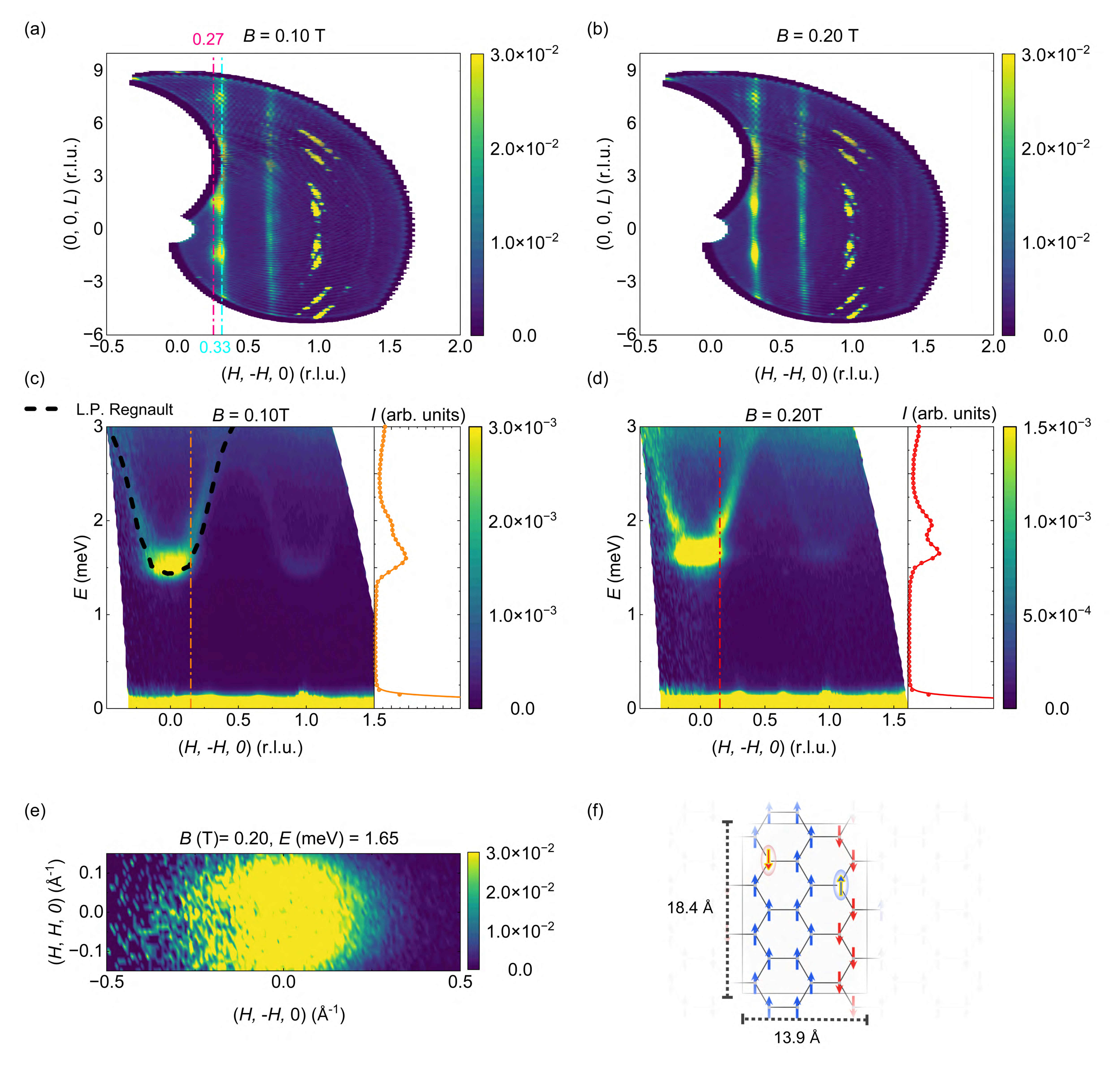}
    \centering
    \caption{\textbf{Field-dependent magnetic structures and excitations.} (a, b) Elastic neutron scattering maps showing a first-order mixed phase (double-zigzag and UUD) at 0.10~T, and a pure UUD state at 0.20~T. (c, d) Corresponding inelastic neutron spectra along $(H,H,0)$. The flat dispersionless band at $\sim$1.65~meV in the UUD state (d) indicates a localized neutral bound state formed by correlated opposite spin-flip excitations. (e) Momentum-space slice of the flat mode at 1.65~meV. (f) Real-space schematic depicting the spatial correlation of the two-magnon bound state.}
    \label{fig:field_dependent_spectra2}
\end{figure}

To determine how the magnetic ground states and associated spin excitations evolve with increasing in-plane magnetic field, we performed neutron scattering measurements at a fixed sample temperature of $\sim$65 mK (Figs.~1a--1c). Figure~2a shows an elastic-scattering reciprocal-space map measured below 0.1 T, where the double-zigzag phase, characterized by an in-plane incommensurate ordering wave vector with $H \sim 0.27$, coexists with UUD order at $H=1/3$. Upon increasing the field to 0.2 T, the UUD Bragg intensity increases at the expense of the incommensurate peak (Fig. 2b), indicating coexistence of the double-zigzag and UUD phases below $\sim$0.3 T. This field range overlaps with the hysteresis observed in magnetization measurements below the one-third magnetization plateau (Fig.~1 and Ref.~\cite{lee25orderbydisorder}).

Figures~2c and 2d show the energy- and momentum-dependent spin excitations measured at 0.1 and 0.2 T, respectively. At 0.1 T, the gapped FM-like spin excitations centered at the $\Gamma$ point exhibit a dispersion consistent with earlier work (black dashed line in Fig.~2c) \cite{REGNAULT2018e00507,doi:10.1073/pnas.2215509119}. At 0.2 T, however, the excitation spectrum separates into two modes (cut along red dashed line in Fig.~2d), inconsistent with the single quadratic dispersion expected for a simple Heisenberg ferromagnet. One branch remains FM-like and can be described by $\Delta_{\mathrm{UUD}}(B,T)+Aq^2$, with an anisotropy gap of $\Delta_{\mathrm{UUD}}\approx 1.7$ meV. In addition, a nearly flat, dispersionless mode appears near 1.65 meV, indicating a localized excitation rather than a collective dispersive spin-wave mode (Fig.~2e). A Fourier transform of its in-plane momentum width gives a characteristic spatial extent of $\sim 14\times18$ \AA$^2$, consistent with dynamic localized UUD clusters (Fig. 2f).

These inelastic results are consistent with the diffraction data at 0.2 T, where the magnetic Bragg peaks associated with UUD order are quasi-two-dimensional and show no long-range coherence along the $c$ axis (Fig.~2b). They are also consistent with earlier reports that found no clear collective spin waves associated with long-range UUD order \cite{10.1016/0378-4363(77)90635-0,REGNAULT1977660,REGNAULT1979194,REGNAULT2006425,REGNAULT2018e00507}.

As the in-plane field $B$ approaches the UUD $\rightarrow$ FM transition, dispersive UUD excitations, the localized mode, and FM spin waves from the field-induced polarized phase coexist. This coexistence coincides with the hysteresis previously reported in magnetization at the transition into the FM state \cite{lee25orderbydisorder}, indicating that the UUD-to-FM transition is first order. Figure~3a shows the field-dependent elastic-scattering map across the UUD-to-FM transition, revealing complete suppression of the quasi-two-dimensional UUD order by 0.7 T.

We performed a detailed field-dependent study across this transition. Below the transition, at 0.45 T, the integrated intensity of the nearly flat mode is comparable to that at 0.2 T but significantly larger than that at 0.1 T (left panel of Fig.~3b and Fig.~3d). At 0.51 T, the flat mode, dispersive UUD excitations, and FM spin waves coexist (center panel of Fig.~3b). By 0.7 T, only FM spin waves remain (right panel of Fig.~3b). Figure~3c summarizes the field dependence of the quasielastic scattering (red box in Fig.~3a), the UUD-associated dispersive spin waves (purple line in the left panel of Fig.~3b), and the localized mode (orange box in the left panel of Fig.~3b).

Throughout the transition region near 0.5 T, the intensities of the $H=1/3$ ordering peak, the UUD-associated dispersive magnon, and the localized excitation all decrease as the intensity of the FM-associated magnon increases (Fig.~3d). The FM magnon has a gap of $\Delta_{\mathrm{FP}}\approx0.75$ meV and exhibits a steeper dispersion, extending up to 3 meV away from $q=0$. In the FM phase for $B>0.59$ T, only this field-polarized magnon is observed, with no remaining signature of the $H=1/3$ ordering peak. Across the measured phase diagram, we find no evidence for a spinon continuum; instead, the spectra are dominated by magnetic order, conventional or localized spin excitations, and phase coexistence. The first-order character of the UUD $\rightarrow$ FM transition is further supported by a lattice distortion inferred from changes in nuclear Bragg intensity \cite{PhysRevB.110.L140407}.

\begin{figure}[h!]
    \includegraphics[width=15cm]{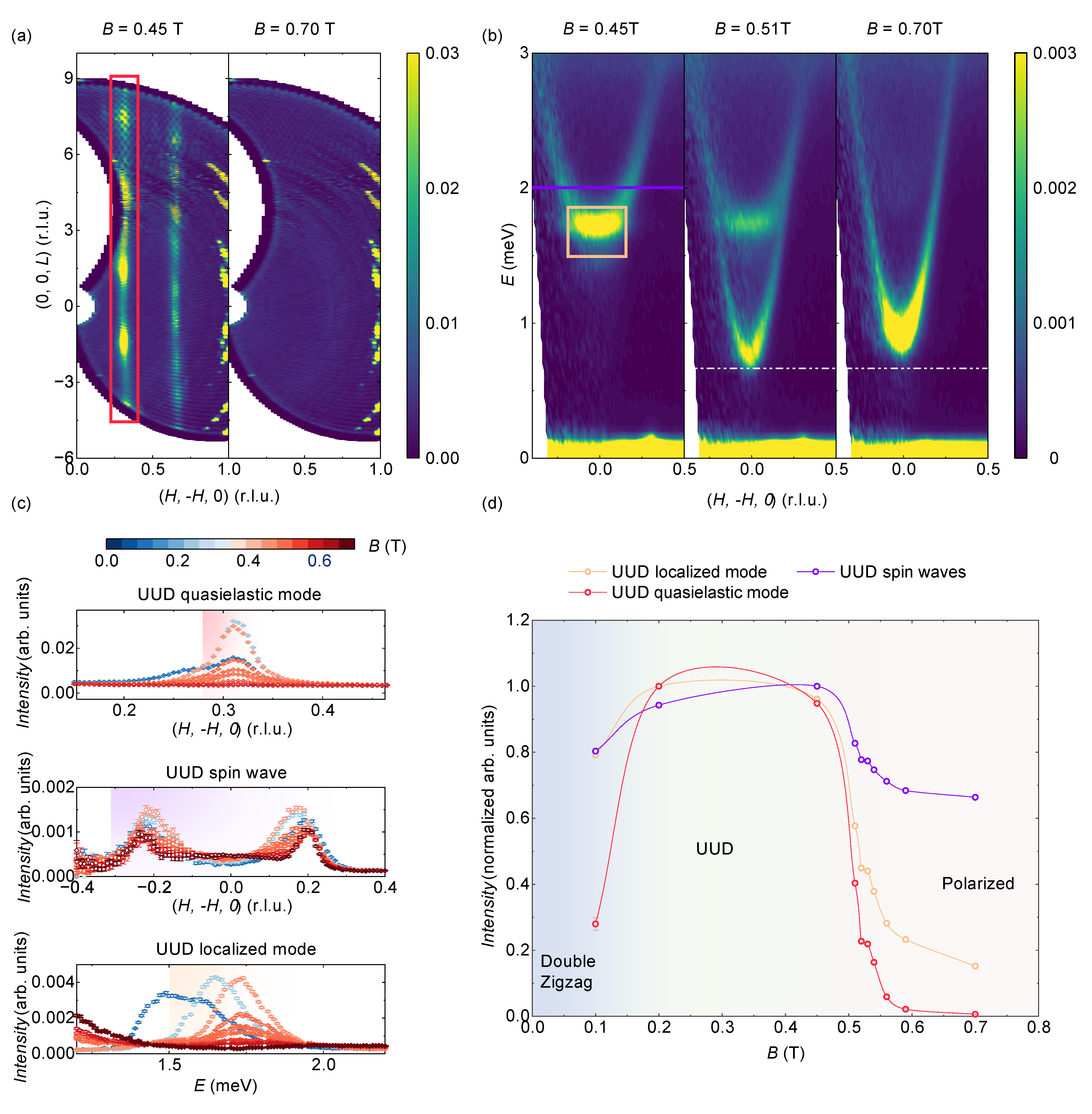}
    \centering
    \caption{\textbf{Collapse of the UUD state and mode evolution driven by magnetic field.} 
     (a) Quasielastic neutron scattering maps at 0.45~T (UUD phase) and 0.70~T (polarized state). Red rectangles highlight the $H \approx 1/3$ magnetic Bragg peaks. (b) Low-energy inelastic spectra along $(H,-H,0)$ exhibiting mode \textcolor{red}{coexistence} near the transition. Purple line and orange box denote integration ranges. (c) Field evolution of the quasielastic peak, UUD dispersive excitation, and localized mode. (d) Normalized integrated intensities versus field. All signatures exhibit a simultaneous collapse at the critical field $B_c \approx 0.54$~T, marking the first-order transition to the polarized state.}
    \label{fig:field_dependent_spectra1}
\end{figure}

Spin transport measurements provide complementary information essential for establishing a full microscopic picture. These measurements require photolithographically patterned devices fabricated on single-crystal surfaces. Plate-like BCAO single crystals were embedded in epoxy and polished with successively finer abrasive media until the exposed $ab$ plane reached a surface roughness below 10 nm (Fig.~1b). The crystallographic orientation was determined by Laue diffraction. Details of the device fabrication are provided in Methods.

Two device geometries were studied. Devices 1 (BCAO.14) and 3 (BCAO.17, see Supporting Information) were configured for local SSE measurements, while device 2 (BCAO.18) was designed for SMR measurements. The SSE data presented below are from device 1. For the SSE measurements, an ac current at 7.7 Hz was applied to the heater, and the inverse spin Hall voltage at both $\omega$ and $2\omega$ was measured on the Pt detector using a lock-in amplifier. Device 2 consisted of two Pt wires, each 10 nm thick, 2 $\mu$m wide, and 900 $\mu$m long, separated laterally by 2 $\mu$m. The resistance of these wires was measured using a home-built Kelvin double bridge at 7.7 Hz as a function of magnetic field within the $ab$ plane, with a sample rotator used to vary the field orientation.

In the SMR \cite{chen16smr,zhang19smr}, thanks to the spin Hall effect (SHE) \cite{dyakonov71she,hirsch99she,sinova15she}, the current in the Pt wire (directed along $x$) produces a spin current of $y$-directed spins along $z$, impinging on the Pt/BCAO interface.  Local exchange processes, encapsulated in a spin mixing conductance $G_{\uparrow \downarrow}$, determine the extent to which spins transfer angular momentum to the BCAO, and conversely the degree to which spins are reflected or accumulate in the Pt.  The inverse spin Hall effect (ISHE) then transduces this to a correction to the resistance of the Pt.  Spin transmission to the magnetic insulator depends in detail on the relative orientation of the spin current moments and the magnetization of the insulator, and it has been argued~\cite{geprags20smrafm} that in multisublattice systems the contributions from sublattices should be considered individually.  

The SMR for Wire 1 ($R = 17.3$~k$\Omega$) on Device 2 at $T = 2$~K is shown in Fig. 4a, b.  (Additional data at higher temperatures are in the Supporting Information.  SMR measurements on  Wire 2 on device 2 and the Pt wire on device 1 are qualitatively identical.)  These data are consistent with the INS results described above, with sharp changes in the MR corresponding with the phase boundaries, and hysteresis observed in magnetization measurements \cite{lee25orderbydisorder}.  This indicates that the magnetization in BCAO at the interface with Pt reflects the bulk response of the crystal.  Hysteresis is clearly observed at the transition between the double zigzag and the UUD phase, and at the transition between the UUD and FM phases.  This is consistent with both field-driven transitions being first-order and exhibiting phase coexistence as seen in the INS measurements.   

\begin{figure}[h!]
    \includegraphics[width=15cm]{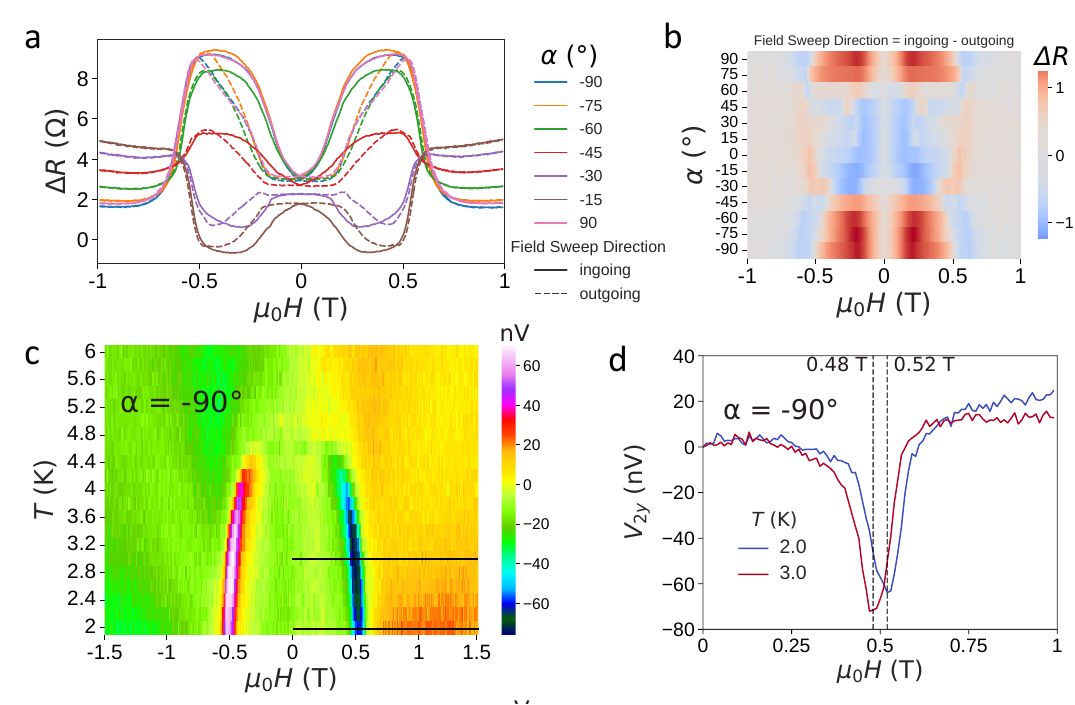}
    \centering
    \caption{Spin transport results on BCAO single crystals.  (a) Spin Hall magnetoresistance traces as a function of in-plane field magnitude and orientation at 2~K, with $\alpha$ defined as in Fig. 1.  Hysteresis is readily apparent, as is the larger SMR magnitude in the UUD phase than the high field FP phase. (b) SMR hysteresis clearly identified through the difference in SMR traces between increasing and decreasing field sweeps. (c) Local SSE response as a function of field and temperature for field oriented transverse to the Pt wire to maximize the SSE signal. A sharp feature is readily apparent in what INS identifies as the UUD/FP coexistence region.  (d) Line-cuts at 2~K and 3~K, showing that the SSE feature has the opposite sign as the SSE in the FP phase.  } 
\end{figure}

In the FM phase, the SMR angular dependence is as expected for a single magnetization fully aligned with the external field, showing a sinusoidal dependence $\propto 1-m_{y}^{2}$, where $m_{y}$  is the component of the magnetization transverse to the Pt wire (see Fig. 1b).  In the UUD phase, the angular dependence of the SMR is more complicated and is a subject for future study. Interestingly, the magnitude of the SMR correction is $\sim 3\times$ larger in magnitude than that in the FM state, implying a larger $G_{\uparrow \downarrow}$  for the UUD phase.  No sharp features are seen in the SMR in the UUD/FM coexistence region, consistent with the smooth evolution of the INS and the magnetization.

The SSE response for device 1 is shown in Figs.~4c and 4d. In magnetic insulators, the SSE can arise from two related mechanisms \cite{jimenezcavero21sse}. First, an interfacial temperature difference between conduction-electron spins in Pt and spin excitations in the magnetic insulator can drive a net spin current across the interface, producing a detectable ISHE voltage in Pt. Second, a temperature gradient within the magnetic insulator can drive a net flux of spin-carrying excitations toward the cold side, either directly or through phonon--spin-excitation scattering \cite{rodriguezsuarez23sse}. In both cases, interfacial exchange, characterized by the spin-mixing conductance $G_{\uparrow\downarrow}$, converts the spin accumulation or spin-excitation flux into a net spin current injected into Pt, generating an ISHE voltage proportional to the heater power.

The sign of the SSE response is determined by the spin Hall angle of Pt and by the orientation of the magnetization carried by the excitations in the insulator that couple to the Pt electron spins. The signal magnitude is maximized when the excitation magnetization, Pt-wire direction, and temperature gradient are mutually orthogonal. The data in Figs.~4c and 4d were obtained with a heater power of 0.2 mW. The continued evolution of the response as the cryostat temperature is lowered indicates that self-heating does not limit the measurement. In the field-induced FM phase, the sign and angular dependence of the SSE are consistent with those observed in Pt on the ferrimagnetic insulator yttrium iron garnet (YIG), where magnons carry magnetization opposite to the bulk magnetization. If the spin-carrying excitations are gapped, the SSE response is suppressed at temperatures well below the gap energy. The small SSE response observed in the double-zigzag, UUD, and FM phases is therefore consistent with dispersive spin excitations whose gaps are comparable to or larger than the relevant thermal energy scale.

In contrast, a sharp SSE feature with a sign opposite to that of the FM-phase signal appears in the UUD/FM coexistence region. This sign reversal indicates thermally driven spin-carrying excitations whose magnetization is aligned with the applied field, opposite to the magnetization carried by conventional magnons in the FM state. The absence of strong gap-induced suppression further suggests that these excitations remain thermally accessible in the coexistence regime. This feature vanishes when the external field is applied parallel to the Pt wire, consistent with the expected SSE symmetry for collinear excitation magnetization.

The sign of the SSE in the coexistence region corresponds to excitations carrying magnetization aligned with the external field. In antiferromagnets and ferrimagnets, multiple magnon branches arise from the presence of inequivalent magnetic sublattices \cite{ohnuma13sse}, and the sign of the measured SSE can depend sensitively on the relative coupling between the inverse spin Hall metal and the different sublattice modes \cite{kikkawa23sse,cramer17magnon}. In BCAO, bond-direction-dependent exchange interactions may also strongly modify the angular-momentum content of dispersive spin excitations. Another possibility is that the SSE is generated by non-magnon spin-carrying excitations whose magnetization is parallel to the applied field, as proposed in prior experiments on materials believed to host triplons \cite{chen21triplon} or one-dimensional gapless spinons \cite{hirobe17spinon}.
However, the possibility of spinons is extremely remote, as the neutron scattering shows magnetic Bragg peaks throughout the phase diagram and no evidence a diffuse continuum of spin excitations in INS that would be expected for such fractionalized excitations\cite{gao2025neutron}.

\section{Theoretical Discussion}

Existing microscopic descriptions of BCAO fall broadly into two classes: models based on weakly bond-dependent XXZ $J_1$--$J_3$ exchange, and models in which strongly bond-direction-dependent, Kitaev-like interactions play a dominant role. We systematically tested representative Hamiltonians from both classes against our combined INS and spin-transport data. Details of this comparison, including the criteria used to identify candidate parameter sets for the two model classes, are provided in the Supporting Information. This comparison favors a model with strongly bond-dependent nearest-neighbor (NN) XXZ exchange, bond-independent third-neighbor (3NN) XXZ exchange, and a weaker isotropic second-neighbor (NNN) interaction. We find that this minimal model simultaneously reproduces several key experimental observations, including: (i) the distinct mode gaps and dispersions in the UUD and field-polarized phases, (ii) the localized two-spin excitation observed by neutron scattering, and (iii) the field-dependent angular-momentum texture of the lowest magnon band, which accounts for the sign of the spin-Seebeck response.  The corresponding Hamiltonian reads as 
\begin{equation}
    \begin{split}
    \mathcal{H} = &\sum_{\langle i,j\rangle\in \gamma} \mathbf{S}^T_i \mathcal{H}^{(1)}_{XXZ,\gamma}\mathbf{S}_j + \!\!\sum_{\langle\!\langle\!\langle i,j\rangle\!\rangle\!\rangle} \mathbf{S}^T_i \mathcal{H}^{(3)}_{XXZ}\mathbf{S}_j \\
    &+ \!\!\sum_{\langle\!\langle i,j\rangle\!\rangle} \mathbf{S}^T_i \mathcal{H}^{(2)}\mathbf{S}_j - g_{xy} \mu_\text{B} \mu_0 H \sum_i S^x_i,
    \end{split}
    \label{FullHam}
\end{equation}
where $\mathbf{S}_i=(S^x_i,S^y_i,S^z_i)$ is the spin at the $i$th site, $g_{xy}=5.1$ the in-plane g factor of the material \cite{doi:10.1073/pnas.2215509119}, $\mu_\text{B}$ the Bohr magneton, $\mu_0 H$  the applied magnetic field,  and $\gamma \in \{x,y,z\}$ labels the three nonequivalent NN bonds of the honeycomb lattice. The NN $z$-bond Hamiltonian takes the form
\begin{equation}
    \mathcal{H}^{(1)}_{XXZ,z} = 
    \begin{pmatrix}
        J^{(1)}_{xy} + D & E & F \\
        E & J^{(1)}_{xy} - D & G \\
        F & G & J^{(1)}_z \\
    \end{pmatrix},
    \label{zbondham}
\end{equation}
with fitted parameters $J^{(1)}_{xy} = -5.63~\text{meV},~J^{(1)}_{z} = -1.64~\text{meV}~~D = 2.46~\text{meV}~E = 0.308~\text{meV}~F = 0~\text{meV}~G = -2.87~\text{meV}$. Because the off-diagonal anisotropies are related by the bond rotations used to generate the $x$ and $y$ bonds, we fix the convention by setting $F=0$ and retaining $G$, following Ref.~\cite{maksimov25kitaev}. The $x$-bond and $y$-bond Hamiltonians are obtained from  \eqref{zbondham} via unitary rotation $U_{\pm 2\pi/3}$ about the $\hat{z}$ axis \cite{doi:10.1073/pnas.2215509119},
while the non-bond-dependent 3NN Hamiltonian reads as
\begin{equation}
    \mathcal{H}^{(3)}_{XXZ} = 
    \begin{pmatrix}
        J^{(3)}_{xy} & 0 & 0 \\
        0 & J^{(3)}_{xy} & 0 \\
        0 & 0 & J^{(3)}_z \\
    \end{pmatrix},
    \label{3NNHam}
\end{equation}
with fitted parameters $J^{(3)}_{xy} = 1.75~\text{meV}$ and $J^{(3)}_{z} = -0.74~\text{meV}$.
Although the second-neighbor exchange is substantially weaker and is often neglected, including it improves the stability of the fitted Hamiltonian and is consistent with previous modeling of BCAO \cite{maksimov25kitaev}. The second-neighbor interaction is isotropic, $\mathcal{H}^{(2)}=J^{(2)}I$, with $J^{(2)}=-0.343~\text{meV}$.
The enhanced bond-dependent parameters $D$ and $E$, together with comparatively small further-neighbor couplings, place BCAO in a strongly bond-anisotropic regime with pronounced Kitaev-like character \cite{maksimov25kitaev,devillez2025bonddependent}, while remaining distinct from the pure Kitaev limit. The resulting excitation spectra and angular-momentum textures are shown in Fig.~5. In the field-polarized phase, the calculated modes correspond to conventional magnons of the long-range-ordered state. By contrast, the finite magnetic correlation length of the UUD phase motivates a description in terms of a locally ordered UUD texture. Accordingly, throughout the following discussion, the terms \emph{magnon} and \emph{spin-wave} refer generically to low-energy excitations near the crystal $\Gamma$ point; in the UUD phase, they should be viewed as excitations of a locally ordered texture rather than long-wavelength collective modes associated with true long-range order at $Q=(1/3,0,0)$.

A central feature of the Hamiltonian~(\ref{FullHam}) is the absence of continuous spin-rotation symmetry about the applied-field direction. As a result, the total spin angular momentum $S^{x}_{\mathrm{tot}}$ is not conserved, and the eigenmodes need not carry a definite $\pm\hbar$ angular momentum. 
Instead, opposite-angular-momentum sectors are generically mixed, so that the angular-momentum content (and hence magnetization content) of a mode becomes a momentum-dependent quantity, $M^x_{n}(\mathbf{k})$, which can even change sign across the Brillouin zone. In the fully polarized phase, however, the absence of a $+\hbar$ sector suppresses such mixing, and the remaining modes become pure spin-angular-momentum eigenstates with well-defined $-\hbar$ character.

Spin-wave spectra calculated using \textsc{SpinW} for the UUD and fully polarized phases reproduce the experimentally observed gaps of the lowest branches at the corresponding fields (Fig.~5). 
The single-magnon spectrum contains no feature that can account for the nearly dispersionless excitation observed in the UUD phase,
 pointing instead to a composite excitation \cite{Sheng2025TwoMagnon} with approximately vanishing total angular momentum, $M^x_{\mathrm{tot}}\simeq0$. Such states arise naturally from pairs of magnons in the lowest UUD band, whose sign-changing angular-momentum texture $M^x_{1}(\mathbf{k})$ allows low-energy combinations with $M^x_{\mathrm{tot}}\simeq0$.
 Bond anisotropy together with the Brillouin-zone folding associated with the six-sublattice magnetic texture further enhances the angular-momentum mixing of the lowest band, thereby enlarging the phase space of such $M^x_{\mathrm{tot}}\approx0$ two-magnon states. Consequently, the lower edge of the corresponding two-magnon continuum is expected at $2\Delta_{\mathrm{min}}\simeq2.56~\mathrm{meV}$, where $\Delta_{\mathrm{min}}$ is the gap of the $M^x_{1}=0$ sector.

This low-energy continuum provides the natural reference for the nearly dispersionless UUD excitation, which we identify as a two-magnon bound state \cite{Bethe1931TwoMag,Wortis1963TwoMag,Torrance1969LocalTwoMag}. To determine the binding scale, we approximate the bound state by a localized, magnetically compensated pair of spin flips embedded in the UUD background.

The binding energy is defined as the energy gain associated with forming a correlated two-spin-flip state relative to two well-separated spin-flip defects \cite{Xie2023SpinFlip}. The strong nearest-neighbor ferromagnetic exchange $J^{(1)}_{xy}$ prevents the formation of a stable bound state on nearest-neighbor bonds. Among all local two-spin-flip configurations, only the third-neighbor configuration possesses a positive binding energy, i.e., 
$E_\text{bind}=4S^2J^{(3)}_{xy}=1.75~\mathrm{meV}$.

The corresponding bound-state energy is then estimated as
$E_\text{bound}=E_{2\mathrm{m}}-\frac{1}{2}E_\text{bind}$. 
Including the weak field dependence of this bound state, as detailed in Sec.~SII of the Supporting Information, yields $E_{\mathrm{bound}} = 1.65~\mathrm{meV}$ at $H = 0.2~\mathrm{T}$, in excellent agreement with the measured energy gap of the nearly dispersionless INS mode.

The field dependence of the spectral weight of the nearly dispersionless mode is naturally explained by its interpretation as a localized two-magnon bound state. Because this excitation can exist only within locally ordered UUD regions, the progressive conversion of UUD into FP regions with increasing field continuously reduces the volume supporting the bound state. Its spectral weight therefore decreases and eventually disappears, precisely as observed by INS. As detailed in Sec.~S3 of the Supporting Information, suppression of this spin-neutral bound-state channel restores spectral weight to spin-carrying single-magnon excitations, contributing to the enhanced spin-Seebeck response near the transition.

A key experimental observation is that the absolute magnitude of the SSE is larger in the UUD phase than in the fully polarized phase (Fig.~4), even though the UUD spectrum is more strongly gapped and therefore supports a lower thermal magnon population. This discrepancy cannot be reconciled within a purely magnonic picture and instead points to an additional channel for angular-momentum transport.  A natural mechanism is provided by the strong magnon-phonon hybridization enabled by weak noncollinearity in the UUD phase \cite{maksimov25kitaev}, allowing acoustic phonons to inherit the angular momentum of the magnetic excitations \cite{Weissenhofer2023RotInv,Zhang2014EdH,Kim2023chiralPhonon,Juraschek2025ChiralPhonons} and sustain a strong low-$T$ spin-Seebeck response \cite{Kikkawa2016SSE,Flebus2014MagnonPolaron}.

\begin{figure}[h!]
    \includegraphics[width=10cm]{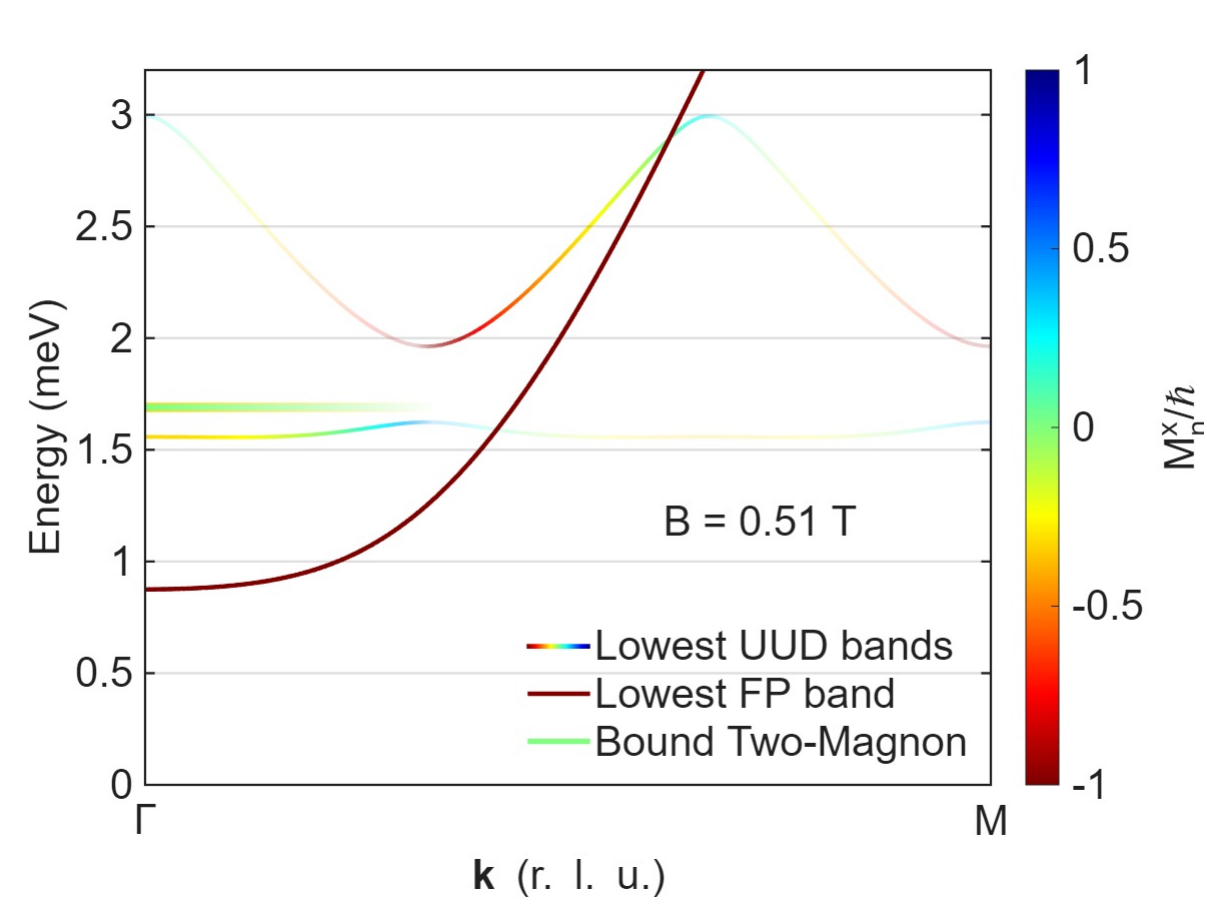}
    \centering
    \caption{Short-wavelength excitation spectrum along $\Gamma$--$M$ in the coexistence regime ($B=0.51$ T) calculated using SpinW. The two lowest UUD modes and the fully polarized mode are shown together with the nearly flat excitation observed by INS, which we assign to a bound two-magnon state. The color scale denotes the angular-momentum content of each mode and the opacity of the bands has been adjusted, as a visual guide, to indicate expected INS visibility. Since true long-range order is absent in the UUD phase, the plotted momentum range is restricted to the short-wavelength sector and avoids the vicinity of the UUD ordering vector. }
    \label{Fig:bands}
\end{figure}

Recent theoretical work \cite{Weissenhofer2025ChiralPhonons,Rafael2023DynPara} has established that efficient angular-momentum transfer through magnon-phonon hybridization can occur also away from magnon-phonon tangent crossing, provided that the spin-lattice interaction is sufficiently strong. BCAO naturally fulfills this condition in the UUD phase, where weak noncollinearity activates a linear exchange-striction channel that is absent in the collinear field-polarized state~\cite{Toth2016Noncollinear1,Park2016Noncollinear2}. Governed by the large nearest-neighbor isotropic exchange $J_{xy}^{(1)}$, this channel dominates the spin-lattice interaction even for small canting angles, allowing the UUD phase to couple strongly to the lattice through exchange-striction in addition to the weaker anisotropy-striction channel $\propto G \ll J_{xy}^{(1)}$. By contrast, the field-polarized phase is largely restricted to the latter.
The resulting effect of strong SLC coupling on transport is further enhanced by the active phonon channels within the two phases. The exchange-striction channel in the canted UUD phase couples primarily to in-plane acoustic strain modes, which have larger velocities than the out-of-plane ZA-like modes associated with the weaker anisotropy-striction channel \cite{doi:10.1073/pnas.2215509119}. Once hybridized, the resulting phonon-like quasiparticles inherit spin angular momentum from the magnetic sector while retaining the larger velocity and thermal occupation of the acoustic phonons. This provides a natural mechanism for the larger absolute spin-Seebeck response observed in the UUD phase.
  It is worth noting that, since the bound state lies close to the single-magnon bands near $\Gamma$, it can also hybridize with the phonon sector. Unlike the single-magnon channel, however, the bound state carries approximately zero net angular momentum, and thus does not contribute to a net spin current.

The pronounced field dependence of the spin-Seebeck response follows from two complementary effects. First, the positive angular-momentum content of the lowest UUD magnon band increases continuously as the field approaches the UUD--FP transition, enhancing the angular momentum transferred to the hybrid phonon branch. Second, the localized two-magnon bound state acts as a field-dependent decay channel for the dispersive spin-carrying excitation \cite{Chernyshev2006NcAFM}. In the UUD phase, cubic magnon interactions couple the one- and two-magnon sectors, allowing the localized bound state to enter the one-magnon self-energy and renormalize the dispersion, linewidth, and coherent spectral weight of the dispersive branch. Since this branch provides the primary channel for magnon-phonon hybridization and angular-momentum transport, the bound state suppresses the spin-Seebeck response by reducing its coherent spectral weight. As the field suppresses the localized bound state, its spectral weight progressively vanishes and this decay channel is removed, restoring the coherent spectral weight of the dispersive excitation and strengthening magnon-phonon hybridization. Together with the increasingly uniform angular-momentum polarization of the lowest magnon band, this naturally accounts for the observed field dependence of the spin-Seebeck response.

\section{Conclusion}

The phase diagram and excitations of BCAO have been controversial for decades.  INS and spin transport measurements in single crystals of BCAO, combined with theoretical calculations, point to a solution. INS observations and hysteresis in SMR demonstrate that phase coexistence takes place at both the double zig-zag/UUD and UUD/FM phase boundaries.  Kitaev-like exchange is critical to stabilizing the UUD and FP phases.  We identify a localized pair defect excitation in the UUD phase responsible for a dispersionless INS signature.  The suppression of the localized pair-defect channel in the UUD/FM coexistence region restores spectral weight to spin-carrying single-magnon excitations and contributes to the enhanced spin-Seebeck response.  The generality of pair defects in Kitaev-like quantum magnets and their impact on thermal transport must be understood when considering candidate observations of QSL signatures in these systems.

Finally, our proposed phonon-mediated contribution to angular-momentum transport is consistent with independent thermodynamic and diffraction studies showing unusually strong spin-lattice coupling in BCAO, manifested by large field-induced changes in the sample length and pronounced anomalies of the nuclear Bragg peaks across the magnetic phase transitions \cite{PhysRevB.110.L140407}. Together, these results indicate that magnetoelastic coupling is not a secondary perturbation but an essential component of the low-energy physics of BCAO, and suggest that similar effects should be considered more broadly in candidate Kitaev magnets.

\section*{Acknowledgments} 
{The single-crystal synthesis and neutron scattering experiments at Rice were supported by the U.S. DOE, BES under Grant Nos. DE-SC0012311 and DE-SC0026179 (P.D). Part of the materials characterization efforts at Rice is supported by the Robert A. Welch Foundation Grant No. C-1839 (P.D.).  The spin transport measurements were supported by NSF DMR-2102028 (T.J.L., D.N.), with resistance bridge development supported by NSF DMR-2329111 (G.E., D.N.). B.F. was supported by  NSF under Grant No. NSF DMR-2144086. R.-Z.L. and C.-L.H. were supported by NSTC under Grant No. NSTC 115-2112-M-006-005. 
The neutron scattering experiments were supported by the Australian Centre for Neutron Scattering, ANSTO, through proposals P15574 and P20132.}

\section{Methods}

\subsection*{Devices and Fabrication}
We present data on two devices fabricated on the $ab$ surface of two different single crystals of BCAO. Device 1, shown in Fig. 1, has three layers. The first is a $10 \; \mathrm{nm}$ by $10 \; \mathrm{\mu m}$ by $800 \; \mathrm{\mu m}$ Pt wire which acts as the spin detector. The second layer is an insulator, SiO$_2$, electrically separating the Pt from the third layer, which is a $50 \; \mathrm{nm}$ by $10 \; \mathrm{\mu m}$ by $1 \; \mathrm{mm}$ wire made of Au which acts as a heater for SSE measurements. This three-layer configuration allows SSE measurements to be performed with the dominant temperature gradient normal to the $ab$ plane, which simplifies interpretation. 

Device 2, comprises two Pt wires roughly $900 \; \mathrm{\mu m}$ long and $10 \; \mathrm{nm}$ thick.  The wires are $2 \; \mathrm{\mu m}$ wide each and $2 \; \mathrm{\mu m}$ apart. For both devices the Pt wires are oriented parallel to the $a - b$ direction.  An image of this device is shown in the Supporting Information.

Bulk crystals must be polished to enable device fabrication. We embed a single crystal of BCAO in a small puck of epoxy, StyCast 1266 for device 1 or StyCast 2850FT for devices 2 and 3, the former is more likely to crack crystals due to shrinkage during cure. We attach the sample to an Allied High Tech brand MultiPrep 8" Precision Polishing System using quick melt wax. We first expose the crystal from the epoxy using $800$ grit sand paper, then polish using 3M lapping paper with progressively smaller particle sizes: $30 \; \mathrm{\mu m}$, then $12 \; \mathrm{\mu m}$, then $3 \; \mathrm{\mu m}$, then $1 \; \mathrm{\mu m}$. Each step is done for 30 seconds to 5 minutes, with progress measured using a stylus profilometer. The final stage is 5 minutes to 30 minutes of polishing with $1 \; \mathrm{\mu m}$ diamond suspension. This often gives a finish with rms roughness below $10 \; \mathrm{nm}$.

Following polishing we perform Laue diffractometry to verify the polished surface is the $ab$ plane and to determine the direction of the $a$ and $b$ axes within the plane. 

We then fabricate the devices using standard photolithographic techniques. For the Pt layers we spin a double layer of resist of LOR15A at $4000$ rpm for $1$ minute, bake for $5$ minutes at $180$ C, then spin on AZ1512 at $4000$ rpm for $1$ minute, and bake for $1$ minute at $110$ C. We then plasma clean and deposit the $10$ nm of Pt using DC sputtering in an Ar atmosphere at a pressure of $3$ mTorr and a sputtering power of $200$ W. The other layers need not have clean edges and so we default to single layers of S1818 which is spun on at $4000$ rpm for $1$ minute and baked at $100$ C for $1$ minute.  We plasma clean to remove residual organics using Ar for $30$ seconds before deposition of each of the layers. Oxygen plasma is not used to minimize the risk of altering the chemistry of the top layers of the BCAO. 

For the spin Seebeck devices, a Pt detector wire 10~nm in thickness, 10 $\mu$m wide, and 800 $\mu$m in length, with leads for four-terminal measurements (see Fig. 1a) was deposited using double-layer photolithography, DC sputtering, and liftoff.  An insulating layer (SiO$_{2}$, 100~nm thick, e-beam evaporated at 0.2~nm/s) was defined and deposited over the Pt detector.  In a subsequent photolithography step, a gold heater wire (50~nm thick, 10~$\mu$m wide, 1 mm long, e-beam evaporated at 0.02~nm/s) was deposited aligned to the Pt detector.  The spin Seebeck data plotted are the antisymmetrized (with field) $y$-component of the second harmonic. 

We measure the SMR through the resistance of the Pt detector. In Fig. 4a,b we show data with a total variation of roughly $10 \; \mathrm{\Omega}$ for device 2, whose Pt wires have a resistance of $17.3 \; \mathrm{k\Omega}$ at $3$~K, so the variation is 1 part in 1000 and the noise to signal is 1 part in 100,000. This signal-to-noise ratio is achieved using a double Kelvin resistance bridge, with the balance  adjusted using a $0.01 \; \mathrm{\Omega}$ resolution decade resistor. The bridge is measured using standard lock-in techniques at 7.7~Hz with ac excitation current of 3.4~$\mu$A.

\subsection*{Magnetic field conditions}
In the SSE measurements, data were collected with magnetic field swept at 10 or 20 Oe/s, from 0~T to the maximum field (1~T or 1.5~T, depending on the run), then to the opposite polarity and back to zero. The SSE data showed no clear indications of hysteresis beyond any temporal lag between the lock-in amplifier measurement and the sweeping field.  In the neutron scattering experiments, the sample was first zero-field-cooled to the measurement temperature. The magnetic field was then swept upward from 0 T to the target field, eventually up to the highest field of 0.7~T. Thus, all of the field-dependent INS measurements were performed on the field-increasing branch.

\clearpage

\setcounter{figure}{0}
\renewcommand{\thefigure}{S\arabic{figure}}
\setcounter{equation}{0}
\renewcommand{\theequation}{S\arabic{equation}}
\setcounter{table}{0}
\renewcommand{\thetable}{S\arabic{table}}
\setcounter{section}{0}
\renewcommand{\thesection}{S\arabic{section}}

\section*{Supporting Information}

\section*{General discussion}

Inelastic neutron scattering (INS) results indicate a finite magnetic correlation length rather than true long-range order in the UUD phase. For this reason, spin-wave calculations should not be interpreted as a literal long-wavelength theory over the entire Brillouin zone. Instead, we use the locally ordered spin texture as a microscopic reference state for short-wavelength magnon-like excitations. This is sufficient for the present purpose, since the nearly dispersionless mode and the neighboring dispersive features observed in INS occur near the crystal $\Gamma$ point, rather than at the magnetic ordering wavevector. The rest of this supplemental text is written within this context and "magnons" or "spin-waves" should be understood as shorthand for magnon-like, short wavelength excitations.

This Supplemental Material is organized as follows. 
Section~S1 starts from the Hamiltonian defined in Eq.~(1) of the main text and discusses the non-definite angular momentum content of the spin-wave modes.
Section~S2 details the association of the non-dispersive mode seen in INS with a bound two-magnon mode and (i) estimates an excitation of the bound two-magnon state, (ii) explains the weak magnetic field dependence, and (iii) provides a minimal model for understanding the features strong localization.
Section~S3 models the phonon mediated spin current described in the main text.
Section~S4 briefly compares the two candidate spin models and their ability to describe both magnetic phases.
Section~S5 presents additional spin Seebeck data and discussion, while Section S6 shows additional SMR data, this from Device 1 (BCAO.14).  Section S7 contains optical micrographs of the wires of Device 2 (BCAO.18).

\section{U(1) Symmetry Breaking}

We model BCAO using a strongly bond-dependent nearest-neighbor XXZ Hamiltonian, supplemented by bond-independent third-neighbor XXZ exchange, a weaker isotropic second-neighbor exchange, and an in-plane magnetic field applied along $\hat{x}$. The anisotropies present in the system explicitly break the continuous spin-rotation symmetry about the field direction. Consequently, $S_{\mathrm{tot}}^x$ is not conserved, bare spin-wave modes with different $S^x$ character can hybridize, and the resulting Bogoliubov-de~Gennes (BdG) eigenmodes are not constrained to carry spin angular momentum of exactly $\pm\hbar$.
The angular momentum content of the $n$th magnon mode is computed within the bosonic BdG formalism from its normalized eigenvector $\psi_{\mathbf{k},n}$ as
\begin{equation}
M^x_{n}(\mathbf{k})= \hbar
\frac{\psi_{\mathbf{k},n}^{\dagger}\Sigma_z S^x \psi_{\mathbf{k},n}}
{\psi_{\mathbf{k},n}^{\dagger}\Sigma_z\psi_{\mathbf{k},n}},
\label{MagCont}
\end{equation}
where $\Sigma_z=\sigma_z\otimes I_6$ is the bosonic BdG metric, and $S^x$ is the spin operator in the Nambu basis $(a_{\mathbf{k},1},\ldots,a_{\mathbf{k},6},a^\dagger_{-\mathbf{k},1},\ldots,a^\dagger_{-\mathbf{k},6})^{T}$, with $a_{\mathbf{k},i}$ ($a^\dagger_{\mathbf{k},i}$) the annihilation (creation) operator for a magnon on the $i$th sublattice.
As shown in Fig.~(5), in the UUD phase the angular-momentum content~\eqref{MagCont} of the $n$th  magnon band forms a momentum-dependent texture that evolves continuously with magnetic field. By contrast, the fully polarized phase represents the trivial mixing limit, in which all magnon bands retain purely $-\hbar$ spin character despite the anisotropies, because no $+\hbar$ sector remains available for hybridization.

\section{Energetics of the Localized Two-Magnon Bound State}

Here we demonstrate that the nearly dispersionless excitation observed in the UUD phase can be identified as a localized two-magnon bound state formed by pairing magnons from the lowest magnon band with opposite spin angular momentum near the zero crossing of its angular-momentum texture~\cite{Bethe1931TwoMag, Wortis1963TwoMag, Sheng2025TwoMagnon}. We first determine the lowest-energy $S_\text{tot}^x=0$ sector of the two-magnon continuum enabled by the broken $U(1)$ symmetry of the Hamiltonian defined by Eqs.~(1)-(3) of the main text. We then estimate the local attractive interaction responsible for binding using a simple flip-pair construction and show that it quantitatively reproduces both the observed excitation energy and its weak field dependence.

\subsection{Two-Magnon Continuum}

The threshold energy of the $S_{\mathrm{tot}}^{x}=0$ two-magnon continuum is defined as
\begin{equation}
E_{2\mathrm{m}}
=
\min_{n,m,\mathbf{k}_1,\mathbf{k}_2}
\left[
\Omega_n(\mathbf{k}_1)+\Omega_m(\mathbf{k}_2)
\right]
\quad
\text{with}
\quad
M_n^x(\mathbf{k}_1)+M_m^x(\mathbf{k}_2)=0,
\label{continuum_edge}
\end{equation}
where $\Omega_n(\mathbf{k})$ is the energy of the $n$th single-magnon band and $M_n^x(\mathbf{k})$ its spin-angular-momentum expectation value~\eqref{MagCont}.  
In conventional magnets with conserved total spin projection along the magnetic field, the lowest spin-neutral ($S_{\mathrm{tot}}^x=0$) two-magnon continuum necessarily involves magnons from distinct spin branches and therefore lies well above the lowest one-magnon excitation.

In the UUD phase of BCAO, however, the broken $U(1)$ symmetry generates a momentum-dependent angular-momentum texture, allowing the spin-neutral two-magnon continuum to be realized entirely within the lowest magnon band. We define $\Delta_\text{min}$ as the lowest single-magnon energy among states satisfying $M^x_{n}(\mathbf{k})=0$. Although the $S^x_\text{tot}=0$ sector of the two-magnon continuum is not restricted to pairs of magnons with individually vanishing angular momentum, the lowest-energy states satisfying the selection rule in Eq.~\eqref{continuum_edge} are concentrated near momenta where the lowest band has $M^x_{1}(\mathbf{k})=0$, corresponding to a zero-crossing in its angular-momentum texture. 
Together with inversion symmetry, $M^x_{n}(\mathbf{k})=M^x_{n}(-\mathbf{k})$, a zero-crossing of $M^x_{1}(\mathbf{k})$ produces a continuous set of nearby paired states whose spin angular momenta cancel while their total momentum remains close to the crystal $\Gamma$ point.
For the parameters relevant to BCAO, $E_{2\mathrm{m}} = 2\Delta_{\text{min}} = 2.56~\text{meV}$.

\subsection{Bound State Energetics}
To estimate the energy of the localized mode, we approximate the bound state by a localized pair of correlated spin flips embedded in the UUD background~\cite{Torrance1969LocalTwoMag, Fukuhara2013LocalFlip, Xie2023SpinFlip}. The corresponding bound-state energy can be written as
\begin{equation}
    E_{\text{bound}}
    =
    E_{2\mathrm{m}}
    -
     E_{\text{bind}},
    \label{defectEnergy}
\end{equation}
where $E_{\mathrm{bind}}>0$ denotes the binding-energy scale of the localized spin-flip pair. 
This real-space construction should not be interpreted as a literal description of the propagating two-magnon eigenstate. Rather, it provides a simple estimate of the local attractive interaction scale responsible for binding by comparing the energy cost of flipping two spins independently with that of flipping the same two spins as a correlated pair.
The energy change associated with flipping a spin in a static spin background is defined as
\begin{equation}
    \Delta E^{(\gamma)}_{ij}
    =
    -2S^2 H^{(\gamma)}_{xx}\sigma_i\sigma_j',
    \label{Eq:DeltaE}
\end{equation}
where $H^{(\gamma)}_{xx}$ is the $S^xS^x$ component of the exchange matrix on bond $\gamma$ shared by spin $i$ and $j$, and $\sigma_i=\pm1$ and $\sigma_j'=\pm1$ denote, respectively, the initial orientation of the flipped spin and static background spin. The endpoint-resolved classical energy gain is defined as the energy gained by creating a correlated pair of spin flips rather than as two isolated spin flips:
\begin{equation}
    \Delta E
    =
    \Delta E_1+\Delta E_2-\Delta E_{\mathrm{pair}} .
\end{equation}
Here $\Delta E_1$ and $\Delta E_2$ are the total bond-energy changes obtained by applying Eq. \eqref{Eq:DeltaE} separately to each isolated spin flip, with the result summed over all bonds connecting the flipped spin to the static background, while $\Delta E_{\mathrm{pair}}$ is the corresponding total energy change when the two flips are created simultaneously as a pair.

Evaluating the bond-energy contributions from the different exchange interactions shows that neither the nearest-neighbor nor the second-neighbor couplings provide an attractive interaction. Instead, the binding originates entirely from the antiferromagnetic third-neighbor exchange, i.e., 
\begin{equation}
    \Delta E
    =
    4S^2J^{(3)}_{xy}.
    \label{eq:bind_energy}
\end{equation}
The local real-space construction considered here is endpoint-resolved, such that the classical energy gain can be written as the sum of the contributions associated with the two flipped-spin endpoints,
$\Delta E = E_{\mathrm{loc}}^{(1)} + E_{\mathrm{loc}}^{(2)} \simeq 2E_{\mathrm{loc}}$.
Because these are two equivalent endpoint contributions to the same composite excitation, the effective pair interaction entering the bound-state energy is therefore
$E_{\mathrm{bind}}\simeq E_{\mathrm{loc}}=\Delta E/2$.
For the lowest bound-state edge considered here, the equivalence of the two endpoint contributions is additionally supported by the symmetry of the constituent magnons discussed in the previous section: they have the same energy and internal angular-momentum character and opposite momenta.
Thus, for $S=1/2$, Eq.~\eqref{eq:bind_energy} yields an estimated binding energy scale
$E_{\mathrm{bind}}\approx0.875~\mathrm{meV}$, indicating that the bound-state energy is set by the third-neighbor exchange scale. This local attraction pulls the two-magnon state below the continuum, stabilizing the non-dispersive excitation.

\subsection{Magnetic Field Dependence}
The weak field dependence of the localized mode can be understood by comparing its energy in the UUD and fully polarized phases. Since the bound spin-flip pair carries no net magnetization, the first-order Zeeman contribution vanishes. The remaining field dependence is therefore governed by the field-driven competition between the UUD and fully polarized phases.
In the UUD phase, the average magnetization per spin is
$M_{\mathrm{UUD}}\simeq M_{\mathrm{s}}/3$, where
$M_{\mathrm{s}}=g\mu_{\mathrm{B}}S$ denotes the saturation magnetization per spin. The corresponding reduction in magnetization relative to the fully polarized state is
\begin{equation}
    \Delta M
    =
    M_{\mathrm{UUD}}-M_{\mathrm{s}}
    =
    -\frac{2}{3}M_{\mathrm{s}}.
\end{equation}
 We estimate the change in energy of a bound two-spin state as
\begin{equation}
    \Delta E
    =
    2\Delta M\cdot(H-H_c)
    =
    -\frac{4}{3}g\mu_\text{B}S(H-H_c),
    \label{fieldDep}
\end{equation}
where $H_c$ is the field at which the system becomes fully polarized. This correction reduces the binding-energy scale as the system approaches the fully polarized phase, slightly raising the bound-state excitation energy. 
\begin{table}[h]
    \centering
    \setlength{\tabcolsep}{8pt}
    \begin{tabular}{|c  c c|}
    \end{tabular}
    \begin{tabular}{|c|c @{\hspace{0.5cm}}c|}
        \hline
       Magnetic Field & $E_{\mathrm{bound}}$ & INS \\
        \hline
        $H=0.20~\mathrm{T}$ & $1.65~\mathrm{meV}$ & $1.65~\mathrm{meV}$ \\
        \hline
        $H=0.45~\mathrm{T}$ & $1.68~\mathrm{meV}$ & $ 1.70~\mathrm{meV}$ \\
        \hline
    \end{tabular}
    \caption{Comparison between the calculated bound-state energy and the non-dispersive feature observed in INS at two applied fields.}
    \label{tab:bound_state_field_dependence}
\end{table}
The calculated shift is small on the scale of the mode energy, consistent with the nearly field-independent flat mode observed in INS. The comparison in Table~\ref{tab:bound_state_field_dependence} shows that this simple estimate captures both the absolute energy scale and the weak upward drift of the dispersionless feature with increasing field.

\section{Phonon-mediated spin~Seebeck response}

In this Supplementary Section, we propose a minimal microscopic mechanism for the anomalous spin-Seebeck signal based on spin-lattice coupling. Because the acoustic phonon dispersions and microscopic spin-lattice coupling parameters of BCAO are not presently known in sufficient detail, we do not attempt a quantitative calculation of the spin-Seebeck coefficient; instead, we identify the symmetry-allowed magnetoelastic channels and their expected evolution between the experimentally observed magnetic phases. We distinguish two features of the measured response: (i) the enhanced spin-Seebeck signal in the UUD phase relative to the fully polarized phase, and (ii) the pronounced field dependence as the system approaches saturation. We argue that the enhanced response in the UUD phase originates from a symmetry-allowed exchange-striction channel activated by the weak canting of the magnetic texture away from $\hat{x}$, which has been reported by Ref.~\cite{maksimov25kitaev} and is consistent with a highly anisotropic system. By contrast, the strong field dependence is associated with the localized two-magnon bound state discussed in Sec.~S2, whose spectral weight and coupling to the single-magnon sector evolve rapidly as the polarized phase is approached.

\subsection{Spin-Lattice Coupling}
The microscopic spin-lattice coupling is obtained by expanding the exchange matrices to linear order in the ionic displacements \cite{Weissenhofer2023RotInv, Weissenhofer2025ChiralPhonons},
\begin{equation}
    \mathcal{H}_{\mathrm{SLC}}
    \simeq
    \sum_\gamma
    \sum_{ij \in \gamma}
    \sum_{\alpha\beta\mu\nu}
    \mathcal{J}^{\alpha\beta,\mu}_{ij}
    S_i^\alpha S_j^\beta
    R_{ij}^{\nu}
    \partial_{\nu}u_i^{\mu},
    \label{Hslc}
\end{equation}
where $S_i = S(\mathbf{r}_i)$, $\mathcal{J}^{\alpha\beta,\mu}_{ij} = \partial H^{(\gamma)}_{\alpha \beta}/ \partial (\delta u^\mu_{ij})$ for a given bond $\gamma$, $R^\nu_{ij}$ is the equilibrium bond vector of the atoms with components $\nu$, and $u^\mu_i= u^\mu (\mathbf{r}_i)$ is the displacement field with component $\mu$. The gradient expansion follows from $\delta u^\mu_{ij}= u_j^\mu-u_i^\mu \simeq R_{ij}^\nu\,\partial_\nu u^\mu_i$. 
For a noncollinear magnetic texture \cite{Toth2016Noncollinear1,Park2016Noncollinear2}, the leading exchange-striction contribution originates from the diagonal exchange channels ($\alpha=\beta$). Retaining the dominant longitudinal contribution ($\mu=\nu$) yields an effective linear magnetoelastic vertex whose non-magnetic component scales as $\propto
\sin(2\theta)
(\epsilon_{xx}+\epsilon_{yy})$, 
where $\theta$ is the canting angle of the spin texture away from $\hat{x}$.
Consequently, through this term, the magnetic excitations couple predominantly to the in-plane longitudinal and transverse acoustic phonons.

In contrast, the collinear fully polarized phase lacks this exchange-striction channel by symmetry. Linear spin--lattice coupling is therefore restricted to the substantially weaker off-diagonal ($\alpha \neq \beta$) anisotropic exchange interactions. These relativistic terms originate from the broken local bond symmetry associated with the puckered Co$^{2+}$ environment~\cite{doi:10.1073/pnas.2215509119} and are expected to couple primarily to out-of-plane acoustic distortions (ZA-like modes). 
In layered quasi-2D materials, these modes typically possess much smaller group velocities than the in-plane longitudinal and transverse acoustic branches, making them a comparatively inefficient channel for spin transport. The distinct magnetoelastic channels active in the two magnetic phases naturally account for the enhanced spin-Seebeck signal in the UUD phase. Even a small canting angle activates efficient exchange-striction coupling to fast in-plane acoustic phonons, whereas the fully polarized phase remains limited to substantially weaker relativistic couplings involving slower out-of-plane distortions.

\subsection{Chiral phonons and dynamic Hybridization}

Although our calculations predict efficient magnetoelastic coupling near magnon--phonon hybridization, the acoustic phonon dispersion of BCAO remains insufficiently characterized to determine whether the low-energy longitudinal and transverse acoustic branches exhibit tangencies or near-degeneracies with the single-magnon spectrum. Such spectral features, however, are not essential to our conclusions. Once time-reversal symmetry is broken, symmetry permits an additional dynamical contribution to the magnetoelastic interaction \cite{Ren2024Adiabatic,Weissenhofer2025ChiralPhonons}, such that the effective magnon--phonon vertex may be written as
\begin{equation}
    g_{n\lambda}(\omega,\mathbf{k})
    =
    g^{(u)}_{n\lambda}(\mathbf{k})
    +
    g^{(\dot u)}_{n\lambda}(\mathbf{k}),
    \label{chiral}
\end{equation}
where $g^{(u)}_{n\lambda}(\mathbf{k})$ is the ordinary displacement- or strain-mediated vertex, while $g^{(\dot u)}_{n\lambda}(\mathbf{k})$ denotes the velocity-coupled dynamic correction associated with the chiral phonon response. 
Microscopically, such velocity-dependent lattice dynamics naturally arise when the ionic motion is coupled to electronic states with finite Berry curvature. In spin-orbit-coupled materials, the electronic ground state acquires a molecular Berry curvature in the space of ionic coordinates once time-reversal symmetry is broken by magnetic order or an applied magnetic field. Integrating out the electronic degrees of freedom then generates an effective Berry-phase contribution to the ionic action, leading to an antisymmetric, velocity-dependent coupling between lattice displacements \cite{Avron1995HallVisc,Barkeshli2012HallVisc,Coh2023ChiralPhonon}. A formally equivalent dynamical contribution can also originate directly from Lorentz forces acting on moving ions \cite{Flebus2023HallVisc}. 
While this dynamical contribution is symmetry-forbidden in the collinear bipartite antiferromagnets with combined $\mathcal{PT}$ symmetry \cite{Coh2023PT} that have been the primary focus of previous theoretical studies, it is symmetry-allowed in more general magnetic systems, including spin-orbit-coupled materials such as BCAO.

Regardless of its microscopic origin, the velocity-dependent contribution
$g^{(\dot u)}_{n\lambda}$
provides an additional channel for magnon--phonon hybridization beyond the immediate vicinity of band crossings \cite{Rafael2023DynPara}, extending the regime over which magnetoelastic effects can remain significant.

\subsection{Sharp Field Dependent Response}

Single magnons coupling to the bound two–magnon sector through cubic terms generated by longitudinal–transverse mixing in the local spin frame provide a natural explanation for the sharp field dependence of the SSE anomaly. Such terms arise from noncollinearity of the ordered state $\propto J_{xy}^{(1)}\sin(2\theta)$ \cite{Chernyshev2006NcAFM}, and in the present material can also be generated by the off-diagonal anisotropies $\propto G \cos(2\theta)$. Schematically, the single- and bound two-magnon hybridize through the $1/S$ cubic vertices
\begin{equation} \mathcal{H}_\text{cube} = \frac{1}{\sqrt{N}} \sum_{\mathbf{k},\mathbf{q}} \sum_{nml} \left[ \Gamma_{nml} \left(\mathbf{k},\mathbf{q}\right) \alpha_{n,\mathbf{k}}^\dagger \alpha_{m,\mathbf{q}} \alpha_{l,\mathbf{k-q}} + \widetilde{\Gamma}_{nml}\left(\mathbf{k},\mathbf{q}\right) \alpha_{n,\mathbf{k}}^\dagger \alpha_{m,-\mathbf{q}}^\dagger \alpha_{l,\mathbf{k+q}} +\mathrm{h.c.} \right], 
\label{Eq:H3BdG} 
\end{equation}
where $\Gamma_{nml}$ and $\widetilde{\Gamma}_{nml}$ are the normal and anomalous terms, respectively, and contain contributions from both the off-diagonal anisotropies and noncollinearity of the ground state. Due to the sharpness of the bound mode, the effect of the two-magnon channel can be represented by an effective self-energy for the single-magnon propagator. For a single-magnon mode $n$, the dressed retarded Green's function is 
\begin{equation} G^R_n(\omega,\mathbf{k}) = \frac{1}{ \omega - \Omega_n(\mathbf{k}) - \Sigma_{nB}(\omega,\mathbf{k}) }, 
\label{Eq:Gbound}
\end{equation}
with effective self energy correction
\begin{equation} \Sigma_{nB}(\omega,\mathbf{k}) = \frac{ |g_{nB}(\mathbf{k})|^2 }{ \omega - E_B(\mathbf{k}) + i\Gamma_B },
\label{Eq:SEBound} 
\end{equation}
where $g_{nB}$ is the projection of the cubic vertex onto the bound two-magnon wavefunction and $E_B(\mathbf{k})-i\Gamma_B$ is the bound two-magnon pole with energy $E_B(\mathbf{k})$ and damping rate $\Gamma_B$. Because the bound two-magnon has no net angular momentum, any single-magnon that decays into the bound two-magnon sector via Eq. \eqref{Eq:Gbound} is unable to contribute to spin transport. The effect is further enhanced for intermediate fields by the strong efficiency of the decay process near the crystal $\Gamma$ point where the lowest single-magnon band's spin content is largest and the two modes are most resonant. In this strongly coupled regime, spin current is heavily suppressed both from direct single-magnon excitations and hybrid magnon-phonon excitations due to the siphoning of single-magnon spectral weight into the two-magnon sector. As the field increases towards the coexistence regime, the spin content of the single-magnon mode continuously approaches a uniform $+\hbar$, reducing the phase volume of and suppressing the bound state. The loss of this competing channel and subsequent restoration of single-magnon spectral weight effectively enhances the phonon-assisted spin Seebeck response in the coexistence regime.

\subsection{Spin-Seebeck Current} 

In the regime where spin--lattice coupling leads to efficient magnon--phonon hybridization, the relevant carriers of angular momentum are the resulting hybrid magnetoelastic quasiparticles. 
 The spin-Seebeck response can therefore written as~\cite{kikkawa16magnonpolaron,Flebus2014MagnonPolaron,Williams2026ChemPot,Kim2023chiralPhonon}
\begin{equation}
    \mathbf{j}_{J^x}
    =
    \int
    \frac{d^3\mathbf{k}}{(2\pi)^3}
    \sum_n
    \left[
    M^x_{s,n}(\mathbf{k})
    +
    M^x_{p,n}(\mathbf{k})
    \right]
    \left[
    \partial_{\mathbf{k}}\Omega_n(\mathbf{k})
    \right]
    \delta f_n(\mathbf{k})\,,
    \label{hybridCurrent}
\end{equation}
where $\mathbf{v}_n(\mathbf{k})=\partial_{\mathbf{k}}\Omega_n(\mathbf{k})$ is the group velocity of the $n$th hybrid mode, $\delta f_n(\mathbf{k})$ is its nonequilibrium bosonic distribution function, $\tau_n(\mathbf{k})$ the  relaxation time, and $M^x_{s,n}$ and $M^x_{p,n}$ denote the spin and lattice contributions, respectively, to the total spin and phonon angular momentum $J^x_\text{tot}$ carried by the hybrid quasiparticle.
Equation~\eqref{hybridCurrent} is intended to illustrate the microscopic transport channels and their relative evolution between the UUD and fully polarized phases, rather than to provide a quantitative fit to the measured SSE magnitude. Within this framework, strong hybridization allows the phonon-like quasiparticles to dominate the angular-momentum transport, carrying angular momentum inherited from the magnon sector while retaining the larger velocities and occupations of acoustic phonons. The difference between the UUD and fully polarized phases is therefore encoded both in the magnetic angular-momentum weight $M^x_{s,n}(\mathbf{k})$, which reflects the hybridization strength, and in the phonon branch character of the hybrid mode. In the UUD phase, the stronger exchange-striction channel can transfer angular momentum to faster in-plane acoustic modes, whereas the fully polarized phase is restricted primarily to weaker coupling involving slower out-of-plane shear-like modes. Taken together, the same hybrid-quasiparticle picture captures both central features of the spin Seekbeck response: a larger absolute signal in the UUD phase and a field-dependent enhancement as the bound-state hybridization channel disappears near the transition.

\section{Comparison of Models}

For both models, we employed an iterative coordinate-search procedure to optimize candidate parameter sets against several key features of the INS data. efore performing this feature optimization, we imposed the necessary criterion that any candidate parameter set $p$ must produce real spin-wave energies throughout the Brillouin zone for both the UUD and field-polarized (FP) states at their respective fields,
\begin{equation}
    \begin{split}
\omega^\text{UUD}_\nu(\mathbf{k};\mathcal{H}(p),H_\text{UUD}) &\in \mathbb{R}\quad \forall\mathbf{k},\nu\\
&\text{and}\\
\omega^\text{FP}_\nu(\mathbf{k};\mathcal{H}(p),H_\text{FP}) &\in \mathbb{R}\quad \forall \mathbf{k},\nu.
    \end{split}
\end{equation}
Here, $\mathcal{H}(p)$ denotes the trial Hamiltonian with parameter set $p$, and $H_\text{UUD}$ and $H_\text{FP}$ denote fields within the corresponding experimental regimes. For a given model, the same Hamiltonian and parameter set are used for both states, since the field-driven transition is not assumed to modify the underlying exchange parameters. This criterion is only a necessary condition for a physically viable parameter set: the presence of imaginary spin-wave modes is sufficient to exclude that set from further consideration.

We explored a broad range of parameters for the model without bond anisotropy, including values near previously reported parameter sets, but did not identify a set satisfying the above criterion for both states simultaneously. Although real spectra could readily be obtained for the FP state, the corresponding UUD spectrum retained imaginary modes for the parameter ranges examined. We therefore were not able to pursue a detailed fit of this reduced model to the INS features.

This difficulty is consistent with previous studies of BaCo$_2$(AsO$_4$)$_2$ using reduced easy-plane XXZ descriptions. In particular, parameter sets obtained from the field-polarized excitation spectrum were found to reproduce the high-field spin-wave dispersion while failing to yield the experimentally observed low-field magnetic structure within the same Hamiltonian \cite{REGNAULT2018e00507}. More recent work has similarly noted that earlier $J_1$--$J_2$--$J_3$ XXZ parameterizations constrained by high-field INS do not simultaneously reproduce the expected low-field state \cite{maksimov25kitaev}. Our inability to identify a common parameter set satisfying the minimal real-spectrum criterion in both phases is therefore consistent with this previously recognized limitation of the reduced XXZ description.

\section{Additional spin Seebeck data}

\subsection{Spin Seebeck discussion}
The local spin Seebeck measurements were taken in a manner similar to that described in previous publications\cite{Luo2024PRB,Luo2024PRBnoise,Luo2025PRB}. We concentrate on Device 1 (BCAO.14), as shown in the main text in Fig. 4c,d.  Lock-in measurements were used with the heater current driven at 7.7~Hz. The second harmonic (spin Seebeck) signal is linear in heater power at low heater powers (below about 0.3~mW), as expected. The spin Seebeck response at high fields (well into the fully polarized regime) has the $\cos$ dependence on field orientation as expected.  
The first-harmonic signals ($x$ or $y$ phases) are essentially featureless, eliminating the possibility that there is any "leakage" contributing to the SSE signal.  There is no detectable dependence of the SSE response on the magnetic field sweep rate beyond what would be expected from the lag of the lock-in amplifier response.  Data shown in Fig. 4c (and below) were taken at 20 Oe/s sweep rates. 

When dealing with small SSE response at low temperatures, one may be concerned about the Nernst-Ettingshausen response of the Pt detector.  We have previously characterized this effect\cite{Luo2023APL} in Pt devices of the same thickness and deposition procedure, fabricated on magnetically inert substrates (SiO$_{2}$ on Si), with the identical SiO$_x$ insulating layer and heater wire configuration as in these measurements. The ordinary Nernst response is linear in magnetic field and heater power, reaching approximately 3.6~nV/T at 3~K and a heater power of 1~mW. While the thermal boundary resistances may be different in the present work, the actual thermal flux through the Pt (which determines the Nernst response) should be identical. At the heater powers used in this work, any ordinary Nernst contribution would be unresolvably small, consistent with the data.  Anomalous Nernst response would require proximitized magnetism in the Pt.  The observed SMR is inconsistent with any induced magnetic order in the Pt.  

Prior experiments on yttrium iron garnet/gadolinium gallium garnet crystals\cite{Luo2024PRBnoise} in the same device geometry show temperature increases of the Pt on the order of 2~K at 5~mW heater power. In the SiO$_{2}$/Si substrate devices in the same configuration\cite{Luo2023APL} (in which the amorphous SiO$_{2}$ is a worse thermal conductor than single crystal substrates), a similar temperature increase of 1.7~K is observed at a heater power of 1~mW.  For the low heater powers (0.2~mW and below) employed in the SSE measurements, temperature increases of the Pt during the measurement are expected to be on the order of 0.1~K. We reemphasize, the absolute magnitude of the SSE signal is set by multiple factors including Pt/BCAO interface quality; the field dependence of the SSE is the key observation rather than the magnitude.

\subsection{BCAO.17}
Data were also acquired on Device 3 (crystal BCAO.17) (Fig. S1).  Due to a short between the gold heater wire and the Pt detector wire, the heater power was limited to lower values to avoid lock-in amplifier issues from crossover of the first harmonic signal.  Even at the considerably reduced heater power, there is a clear, sharp signature of the SSE response at the transition between UUD and FM states, closely resembling that observed in Device 1 on crystal BCAO.14.  

\begin{figure}[t]
    \includegraphics[width=14cm]{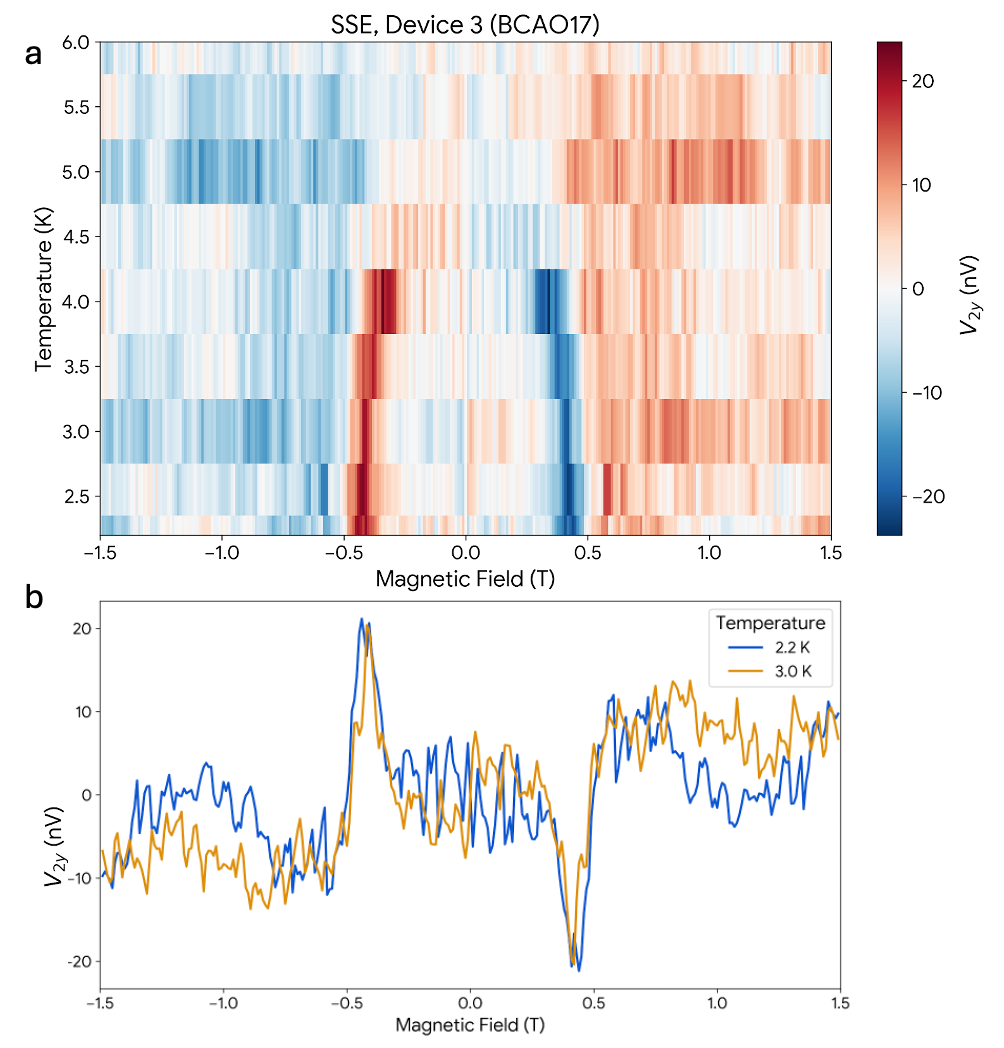}
    \centering
    \caption{Spin Seebeck data on Device 3 (BCAO.17), antisymmetrized second harmonic as a function of temperature and field. Field is oriented along $\alpha = -90^{\circ}$, as in Fig. 4c of the main text.  Heater power was limited to 0.02~mW, limiting signal-to-noise compared with the data on BCAO.14 shown in Fig. 4c of the main text.  Line cuts are shown at 2.2~K and 3~K.} 
\end{figure}

\clearpage

\subsection{BCAO.14, higher temperature}

Additional SSE data on Device 1 (BCAO.14) also extends to higher temperatures.  These are shown in Fig. S2.  

\begin{figure}[h!]
    \includegraphics[width=14cm]{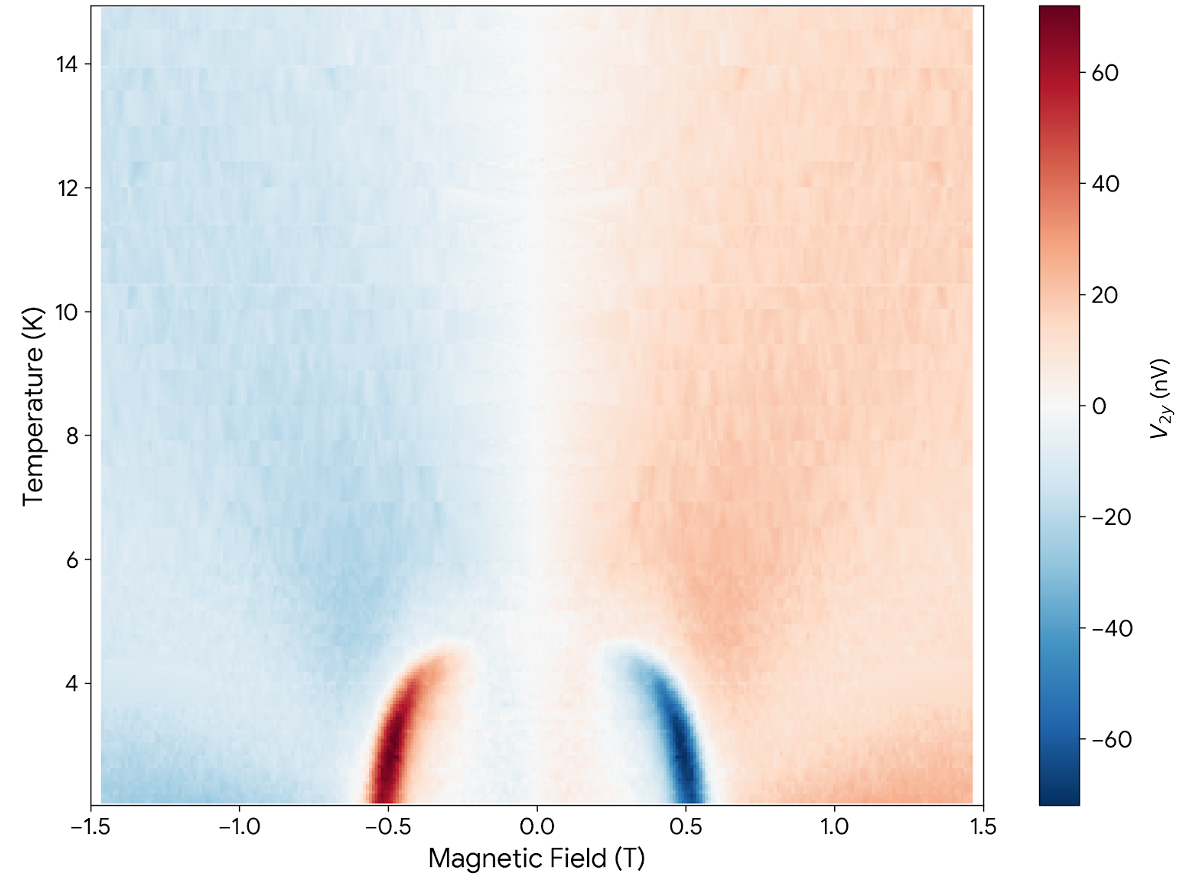}
    \centering
    \caption{Spin Seebeck data on Device 1 (BCAO.14), antisymmetrized second harmonic as a function of temperature and field, 0.2~mW heater power, extended to 15~K. Field is along $\alpha=-90^{\circ}$. } 
\end{figure}

\clearpage

Antisymmetrization as a function of field has been applied to the data in Fig. 4c and in Fig. S2. In practice, all this does is subtract a very small ($\sim$15~nV) offset from the data.  Fig. S3 shows the data from Fig. 4c without the antisymmetrization procedure.  

\begin{figure}[h!]
    \includegraphics[width=8.5cm]{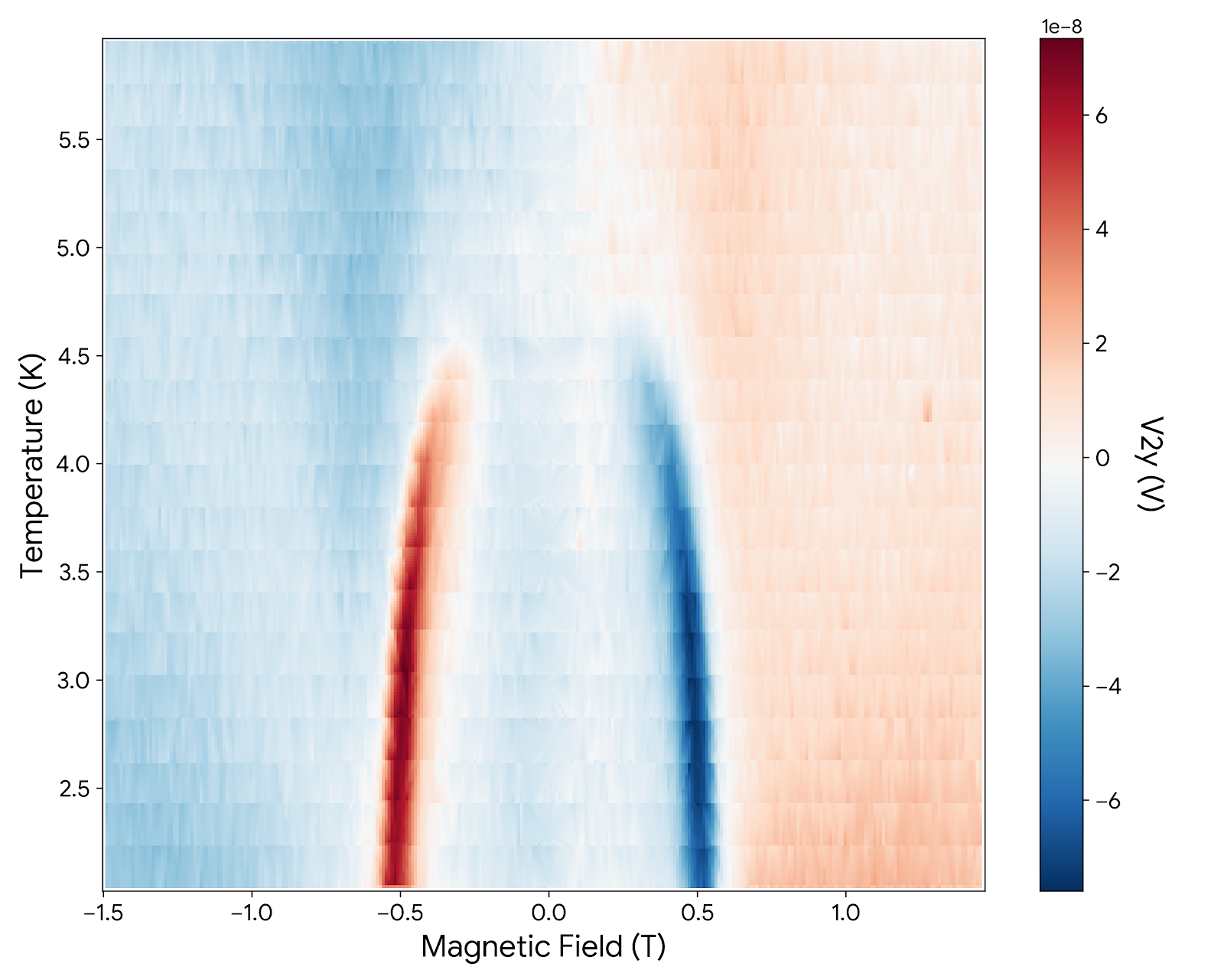}
    \centering
    \caption{Spin Seebeck data on Device 1 (BCAO.14), as above and as in Fig. 4c,d but without the antisymmetrization.} 
\end{figure}

\clearpage

\section{Additional SMR data}

SMR data were also acquired on Device 1 (BCAO.14) at several temperatures ($R \approx 1525$~$\Omega$).  While the details of the SMR curves differ slightly from those on Device 2 Wire 1 (Fig. 4a,b), the qualitative features are the same.  

\begin{figure}[h!]
    \includegraphics[width=\linewidth]{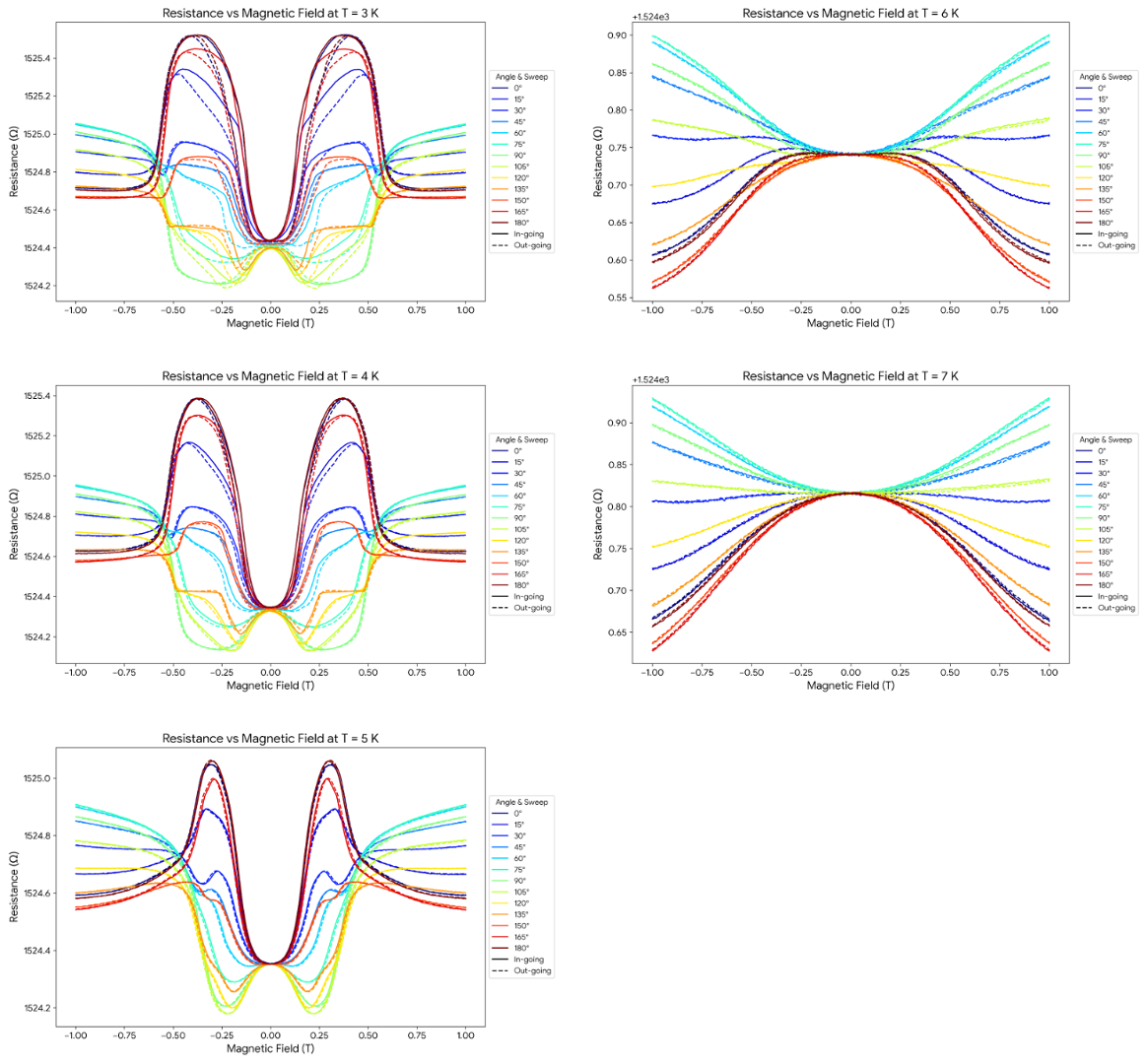}
    \centering
    \caption{SMR data on Device 1 (BCAO.14), measurement current 80~$\mu$A, 7.7~Hz. Angle definitions are shifted by 90 degrees from those in the main text.} 
\end{figure}

\clearpage

\section{BCAO.18 device}

Figure S5 shows optical micrographs of the two wires on Device 2, the SMR device with data in Fig. 4a,b.  

\begin{figure}[h!]
    \includegraphics[width=10cm]{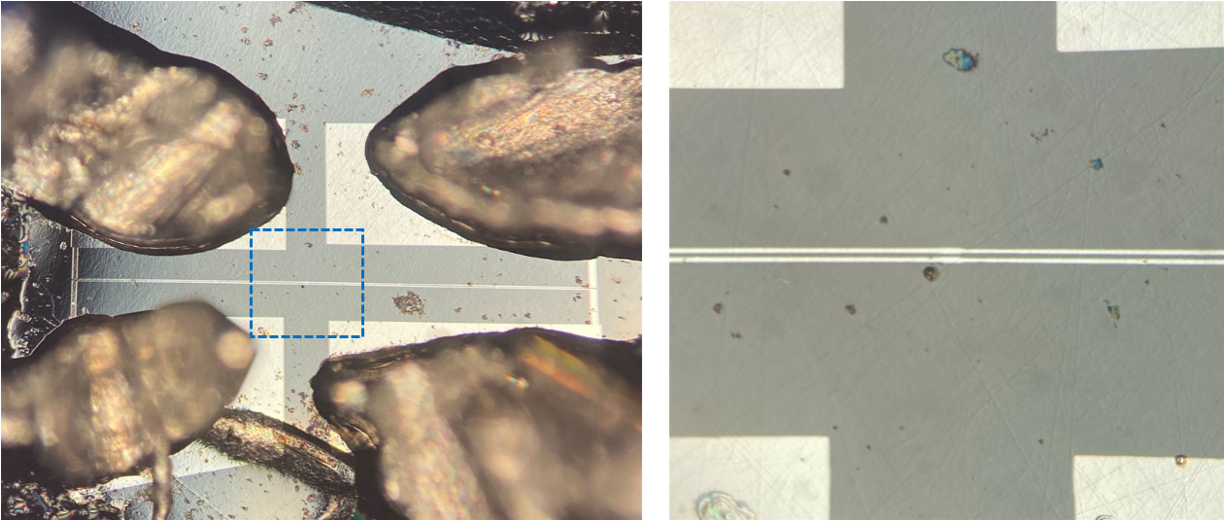}
    \centering
    \caption{The two Pt wires on Device 2.  The wires are 2~$\mu$m wide, separated by 2~$\mu$m, and each are 900~$\mu$m in length. The right panel shows the indicated region of the left field of view.  Measurement wires have been attached to the Pt pads by conductive epoxy.} 
\end{figure}

\clearpage

\bibliography{bibliography}

@article{broholm20qsl,
   author = {Broholm, C. and Cava, R. J. and Kivelson, S. A. and Nocera, D. G. and Norman, M. R. and Senthil, T.},
   title = {Quantum spin liquids},
   journal = {Science},
   volume = {367},
   number = {6475},
   pages = {eaay0668},
   DOI = {10.1126/science.aay0668},
   url = {http://science.sciencemag.org/content/367/6475/eaay0668.abstract},
   year = {2020},
   type = {Journal Article}
}

@article{3m4m-3v59,
  title = {Kitaev quantum spin liquids},
  author = {Matsuda, Yuji and Shibauchi, Takasada and Kee, Hae-Young},
  journal = {Rev. Mod. Phys.},
  volume = {97},
  issue = {4},
  pages = {045003},
  numpages = {61},
  year = {2025},
  month = {Dec},
  publisher = {American Physical Society},
  doi = {10.1103/3m4m-3v59},
  url = {https://link.aps.org/doi/10.1103/3m4m-3v59}
}

@article{Takagi2019KitaevQSL,
  author    = {Hidenori Takagi and Tomohiro Takayama and George Jackeli and Giniyat Khaliullin and Stephen E. Nagler},
  title     = {Concept and realization of Kitaev quantum spin liquids},
  journal   = {Nature Reviews Physics},
  volume    = {1},
  number    = {4},
  pages     = {264--280},
  year      = {2019},
  doi       = {10.1038/s42254-019-0038-2}
}

@article{chen21triplon,
   author = {Chen, Yao and Sato, Masahiro and Tang, Yifei and Shiomi, Yuki and Oyanagi, Koichi and Masuda, Takatsugu and Nambu, Yusuke and Fujita, Masaki and Saitoh, Eiji},
   title = {Triplon current generation in solids},
   journal = {Nat. Comm.},
   volume = {12},
   number = {1},
   pages = {5199},
   ISSN = {2041-1723},
   DOI = {10.1038/s41467-021-25494-7},
   url = {https://doi.org/10.1038/s41467-021-25494-7},
   year = {2021},
   type = {Journal Article}
}

@article{chen16smr,
  author  = {Chen, Yan-Ting and Takahashi, Saburo and Nakayama, Hiroyasu and Althammer, Matthias and Goennenwein, Sebastian T. B. and Saitoh, Eiji and Bauer, Gerrit E. W.},
  title   = {Theory of spin Hall magnetoresistance ({SMR}) and related phenomena},
  journal = {Journal of Physics: Condensed Matter},
  volume  = {28},
  number  = {10},
  pages   = {103004},
  year    = {2016},
  doi     = {10.1088/0953-8984/28/10/103004},
  url     = {https://doi.org/10.1088/0953-8984/28/10/103004}
}

@article{cramer17magnon,
   author = {Cramer, Joel and Guo, Er-Jia and Gepr{\"a}gs, Stephan and Kehlberger, Andreas and Ivanov, Yurii P. and Ganzhorn, Kathrin and Della Coletta, Francesco and Althammer, Matthias and Huebl, Hans and Gross, Rudolf and Kosel, Jürgen and Kläui, Mathias and Goennenwein, Sebastian T. B.},
   title = {Magnon Mode Selective Spin Transport in Compensated Ferrimagnets},
   journal = {Nano Letters},
   volume = {17},
   number = {6},
   pages = {3334-3340},
   ISSN = {1530-6984},
   DOI = {10.1021/acs.nanolett.6b04522},
   url = {https://doi.org/10.1021/acs.nanolett.6b04522},
   year = {2017},
   type = {Journal Article}
}

@article{geprags20smrafm,
   author = {Gepr{\"a}gs, Stephan and Opel, Matthias and Fischer, Johanna and Gomonay, Olena and Schwenke, Philipp and Althammer, Matthias and Huebl, Hans and Gross, Rudolf},
   title = {Spin {Hall} magnetoresistance in antiferromagnetic insulators},
   journal = {J. Appl. Phys.},
   volume = {127},
   number = {24},
   pages = {243902},
   ISSN = {0021-8979},
   DOI = {10.1063/5.0009529},
   url = {https://doi.org/10.1063/5.0009529},
   year = {2020},
   type = {Journal Article}
}

@article{hirobe17spinon,
   author = {Hirobe, Daichi and Sato, Masahiro and Kawamata, Takayuki and Shiomi, Yuki and Uchida, Ken-ichi and Iguchi, Ryo and Koike, Yoji and Maekawa, Sadamichi and Saitoh, Eiji},
   title = {One-dimensional spinon spin currents},
   journal = {Nature Physics},
   volume = {13},
   number = {1},
   pages = {30-34},
   ISSN = {1745-2481},
   DOI = {10.1038/nphys3895},
   url = {https://doi.org/10.1038/nphys3895},
   year = {2017},
   type = {Journal Article}
}

@article{jimenezcavero21sse,
   author = {Jim{\'e}nez-Cavero, P. and Lucas, I. and Bugallo, D. and L{\'o}pez-Bueno, C. and Ramos, R. and Algarabel, P. A. and Ibarra, M. R. and Rivadulla, F. and Morell{\'o}n, L.},
   title = {Quantification of the interfacial and bulk contributions to the longitudinal spin {Seebeck} effect},
   journal = {Appl. Phys. Lett.},
   volume = {118},
   number = {9},
   pages = {092404},
   ISSN = {0003-6951},
   DOI = {10.1063/5.0038192},
   url = {https://doi.org/10.1063/5.0038192},
   year = {2021},
   type = {Journal Article}
}

@article{kato25sse,
   author = {Kato, Yasuyuki and Nasu, Joji and Sato, Masahiro and Okubo, Tsuyoshi and Misawa, Takahiro and Motome, Yukitoshi},
   title = {Spin {Seebeck} Effect as a Probe for {Majorana} Fermions in {Kitaev} Spin Liquids},
   journal = {Phys. Rev. X},
   volume = {15},
   number = {1},
   pages = {011050},
   DOI = {10.1103/PhysRevX.15.011050},
   url = {https://link.aps.org/doi/10.1103/PhysRevX.15.011050},
   year = {2025},
   type = {Journal Article}
}

@article{kikkawa23sse,
   author = {Kikkawa, Takashi and Saitoh, Eiji},
   title = {Spin {Seebeck} Effect: Sensitive Probe for Elementary Excitation, Spin Correlation, Transport, Magnetic Order, and Domains in Solids},
   journal = {Ann. Rev. Cond. Matt. Phys.},
   volume = {14},
   number = {1},
   pages = {129-151},
   ISSN = {1947-5454},
   DOI = {10.1146/annurev-conmatphys-040721-014957},
   url = {https://doi.org/10.1146/annurev-conmatphys-040721-014957},
   year = {2023},
   type = {Journal Article}
}

@article{kikkawa16magnonpolaron,
   author = {Kikkawa, Takashi and Shen, Ka and Flebus, Benedetta and Duine, Rembert A. and Uchida, Ken-ichi and Qiu, Zhiyong and Bauer, Gerrit E. W. and Saitoh, Eiji},
   title = {Magnon Polarons in the Spin {Seebeck} Effect},
   journal = {Phys. Rev. Lett.},
   volume = {117},
   number = {20},
   pages = {207203},
   DOI = {10.1103/PhysRevLett.117.207203},
   url = {https://link.aps.org/doi/10.1103/PhysRevLett.117.207203},
   year = {2016},
   type = {Journal Article}
}

@article{lee25orderbydisorder,
   author = {Lee, Sangyun and Zhang, Shengzhi and Thomas, S. M. and Pressley, L. and Bridges, C. A. and Choi, Eun Sang and Zapf, Vivien S. and Winter, Stephen M. and Lee, Minseong},
   title = {{Quantum order by disorder is a key to understanding the magnetic phases of BaCo$_{2}$(AsO$_{4}$)$_{2}$}},
   journal = {npj Quantum Materials},
   volume = {10},
   number = {1},
   pages = {11},
   ISSN = {2397-4648},
   DOI = {10.1038/s41535-025-00728-9},
   url = {https://doi.org/10.1038/s41535-025-00728-9},
   year = {2025},
   type = {Journal Article}
}

@article{li20thermalxport,
   author = {Li, Mingda and Chen, Gang},
   title = {Thermal transport for probing quantum materials},
   journal = {MRS Bulletin},
   volume = {45},
   number = {5},
   pages = {348-356},
   ISSN = {0883-7694},
   DOI = {10.1557/mrs.2020.124},
   url = {https://www.cambridge.org/core/product/F60114554A9FDA9010F03BD440865EEC},
   year = {2020},
   type = {Journal Article}
}

@article{ohnuma13sse,
   author = {Ohnuma, Yuichi and Adachi, Hiroto and Saitoh, Eiji and Maekawa, Sadamichi},
   title = {Spin {Seebeck} effect in antiferromagnets and compensated ferrimagnets},
   journal = {Physical Review B},
   volume = {87},
   number = {1},
   pages = {014423},
   DOI = {10.1103/PhysRevB.87.014423},
   url = {https://link.aps.org/doi/10.1103/PhysRevB.87.014423},
   year = {2013},
   type = {Journal Article}
}

@article{rodriguezsuarez23sse,
   author = {Rodr{\'i}guez-Su{\'a}rez, Roberto L. and Rezende, Sergio M.},
   title = {Dominance of the phonon drag mechanism in the spin {Seebeck} effect at low temperatures},
   journal = {Phys. Rev. B},
   volume = {108},
   number = {13},
   pages = {134407},
   DOI = {10.1103/PhysRevB.108.134407},
   url = {https://link.aps.org/doi/10.1103/PhysRevB.108.134407},
   year = {2023},
   type = {Journal Article}
}

@article{savary16qsl,
   author = {Savary, Lucile and Balents, Leon},
   title = {Quantum spin liquids: a review},
   journal = {Rep. Prog. Phys.},
   volume = {80},
   number = {1},
   pages = {016502},
   ISSN = {0034-4885
1361-6633},
   DOI = {10.1088/0034-4885/80/1/016502},
   url = {http://dx.doi.org/10.1088/0034-4885/80/1/016502},
   year = {2016},
   type = {Journal Article}
}

@article{tu25bcao,
title = {Evidence for Mobile Gapless Spinons in a Honeycomb Lattice},
journal = {Chin. Phys. Lett.},
volume = {42},
number = {6},
pages = {067304},
year = {2025},
issn = {},
doi = {10.1088/0256-307X/42/6/067304},	
URL = {http://cpl.iphy.ac.cn/en/article/doi/10.1088/0256-307X/42/6/067304},
author = {Chengpeng Tu and Dongzhe Dai and Xu Zhang and Chengcheng Zhao and Xiaobo Jin and Bin Gao and Tong Chen and Pengcheng Dai and Shiyan Li}
}

@article{zhang23bcaothz,
    author  = {Zhang, Xinshu and Xu, Yuanyuan and Halloran, T. and Zhong, Ruidan and Broholm, C. and Cava, R. J. and Drichko, N. and Armitage, N. P.},
   title   = {A magnetic continuum in the cobalt-based honeycomb magnet {BaCo}$_{2}$({AsO}$_{4}$)$_{2}$},
   journal = {Nat. Mater.},
   volume  = {22},
   number  = {1},
   pages   = {58--63},
   DOI     = {10.1038/s41563-022-01403-1},
   url     = {https://doi.org/10.1038/s41563-022-01403-1},
   year    = {2023}
}

@article{REGNAULT2018e00507,
title = {{Polarized-neutron investigation of magnetic ordering and spin dynamics in BaCo$_{2}$(AsO$_{4}$)$_{2}$ frustrated honeycomb-lattice magnet}},
journal = {Heliyon},
volume = {4},
number = {1},
pages = {e00507},
year = {2018},
issn = {2405-8440},
doi = {https://doi.org/10.1016/j.heliyon.2018.e00507},
url = {https://www.sciencedirect.com/science/article/pii/S2405844017332012},
author = {L.-P. Regnault and C. Boullier and J.E. Lorenzo}
}

@article{REGNAULT1979194,
title = {{Effect of a magnetic field on the magnetic ordering of BaCo$_{2}$(AsO$_{4}$)$_{2}$}},
journal = {Journal of Magnetism and Magnetic Materials},
volume = {14},
number = {2},
pages = {194-196},
year = {1979},
issn = {0304-8853},
doi = {https://doi.org/10.1016/0304-8853(79)90117-3},
url = {https://www.sciencedirect.com/science/article/pii/0304885379901173},
author = {L.P. Regnault and J. Rossat-Mignod}
}

@article{doi:10.1126/sciadv.aay6953,
author = {Ruidan Zhong  and Tong Gao  and Nai Phuan Ong  and Robert J. Cava },
title = {Weak-field induced nonmagnetic state in a Co-based honeycomb},
journal = {Science Advances},
volume = {6},
number = {4},
pages = {eaay6953},
year = {2020},
doi = {10.1126/sciadv.aay6953},
URL = {https://www.science.org/doi/abs/10.1126/sciadv.aay6953}}

@article{zhang19smr,
   author = {Zhang, Xian-Peng and Bergeret, F. Sebastian and Golovach, Vitaly N.},
   title = {Theory of Spin {Hall} Magnetoresistance from a Microscopic Perspective},
   journal = {Nano Letters},
   volume = {19},
   number = {9},
   pages = {6330-6337},
   ISSN = {1530-6984},
   DOI = {10.1021/acs.nanolett.9b02459},
   url = {https://doi.org/10.1021/acs.nanolett.9b02459},
   year = {2019},
   type = {Journal Article}
}

@article{zhou17qslrmp,
   author = {Zhou, Yi and Kanoda, Kazushi and Ng, Tai-Kai},
   title = {Quantum spin liquid states},
   journal = {Rev. Mod. Phys.},
   volume = {89},
   number = {2},
   pages = {025003},
   DOI = {10.1103/RevModPhys.89.025003},
   url = {https://link.aps.org/doi/10.1103/RevModPhys.89.025003},
   year = {2017},
   type = {Journal Article}
}

@article{dyakonov71she,
    title = {Possibility of Orienting Electron Spins with Current},
    author = {D'yakonov, M. I. and Perel', V. I.},
    journal = {JETP Lett.},
    volume = {13},
    issue = {11},
    pages = {657},
    year = {1971},
    doi = {},
    url = {http://jetpletters.ru/ps/0/article_24366.shtml},
}

@article{hirsch99she,
  title = {Spin {Hall} Effect},
  author = {Hirsch, J. E.},
  journal = {Phys. Rev. Lett.},
  volume = {83},
  issue = {9},
  pages = {1834--1837},
  numpages = {0},
  year = {1999},
  month = {Aug},
  publisher = {American Physical Society},
  doi = {10.1103/PhysRevLett.83.1834},
  url = {https://link.aps.org/doi/10.1103/PhysRevLett.83.1834}
}

@article{sinova15she,
  title = {Spin {Hall} effects},
  author = {Sinova, Jairo and Valenzuela, Sergio O. and Wunderlich, J. and Back, C. H. and Jungwirth, T.},
  journal = {Rev. Mod. Phys.},
  volume = {87},
  issue = {4},
  pages = {1213--1260},
  numpages = {47},
  year = {2015},
  month = {Oct},
  publisher = {American Physical Society},
  doi = {10.1103/RevModPhys.87.1213},
  url = {https://link.aps.org/doi/10.1103/RevModPhys.87.1213}
}

@article{maksimov25kitaev,
  title = {{Strong Kitaev interaction in ${\text{BaCo}}_{2}({\mathrm{AsO}}_{4}{)}_{2}$}},
  author = {Maksimov, Pavel A. and Jiang, Shengtao and Regnault, L. P. and Chernyshev, A. L.},
  journal = {Phys. Rev. Lett.},
  volume = {135},
  issue = {6},
  pages = {066703},
  numpages = {10},
  year = {2025},
  month = {Aug},
  publisher = {American Physical Society},
  doi = {10.1103/k1gq-k8m7},
  url = {https://link.aps.org/doi/10.1103/k1gq-k8m7}
}

@article{devillez2025bonddependent,
        title = {{Bond-dependent interactions and ill-ordered state in the honeycomb cobaltate ${\mathrm{BaCo}}_{2}{({\mathrm{AsO}}_{4})}_{2}$}},
  author = {Devillez, A. and Robert, J. and Lhotel, E. and Ballou, R. and Cavenel, C. and Denis Romero, F. and Faure, Q. and Jacobsen, H. and Lass, J. and Mazzone, D. G. and Hansen, U. Bengaard and Enderle, M. and Raymond, S. and De Brion, S. and Simonet, V. and Songvilay, M.},
  journal = {Phys. Rev. Res.},
  volume = {7},
  issue = {4},
  pages = {L042040},
  numpages = {7},
  year = {2025},
  month = {Nov},
  publisher = {American Physical Society},
  doi = {10.1103/f9sk-zfqc},
  url = {https://link.aps.org/doi/10.1103/f9sk-zfqc}
}

@article{doi:10.1073/pnas.2215509119,
author = {Thomas Halloran  and Félix Desrochers  and Emily Z. Zhang  and Tong Chen  and Li Ern Chern  and Zhijun Xu  and Barry Winn  and M. Graves-Brook  and M. B. Stone  and Alexander I. Kolesnikov  and Yiming Qiu  and Ruidan Zhong  and Robert Cava  and Yong Baek Kim  and Collin Broholm },
title = {{Geometrical frustration versus Kitaev interactions in BaCo$_{2}$(AsO$_{4}$)$_{2}$}},
journal = {Proceedings of the National Academy of Sciences},
volume = {120},
number = {2},
pages = {e2215509119},
year = {2023},
doi = {10.1073/pnas.2215509119},
URL = {https://www.pnas.org/doi/abs/10.1073/pnas.2215509119}}

@article{PhysRevB.110.L140407,
  title = {Intermediate field-induced phase of the honeycomb magnet ${{\mathrm{BaCo}}}_{2}{({\mathrm{AsO}}_{4})}_{2}$},
  author = {Mukharjee, Prashanta K. and Shen, Bin and Erdmann, Sebastian and Jesche, Anton and Kaiser, Julian and Baral, Priya R. and Zaharko, Oksana and Gegenwart, Philipp and Tsirlin, Alexander A.},
  journal = {Phys. Rev. B},
  volume = {110},
  issue = {14},
  pages = {L140407},
  numpages = {6},
  year = {2024},
  month = {Oct},
  publisher = {American Physical Society},
  doi = {10.1103/PhysRevB.110.L140407},
  url = {https://link.aps.org/doi/10.1103/PhysRevB.110.L140407}
}

@article{REGNAULT1977660,
title = {{Magnetic ordering in a planar X-Y model: BaCo$_{2}$(AsO$_{4}$)$_{2}$}},
journal = {Physica B+C},
volume = {86-88},
pages = {660-662},
year = {1977},
issn = {0378-4363},
doi = {https://doi.org/10.1016/0378-4363(77)90635-0},
url = {https://www.sciencedirect.com/science/article/pii/0378436377906350},
author = {L.P. Regnault and P. Burlet and J. Rossat-Mignod}
}

@article{REGNAULT2006425,
title = {Investigation by spherical neutron polarimetry of magnetic properties in ${\mathrm{BaCo}}_{2}{({\mathrm{AsO}}_{4})}_{2}$},
journal = {Physica B: Condensed Matter},
volume = {385-386},
pages = {425-427},
year = {2006},
issn = {0921-4526},
doi = {https://doi.org/10.1016/j.physb.2006.05.142},
url = {https://www.sciencedirect.com/science/article/pii/S0921452606010209},
author = {L.P. Regnault and C. Boullier and J.Y. Henry}
}

@article{Sheng2025TwoMagnon,
  author  = {Sheng, Jieming and Mei, Jia-Wei and Wang, Le and Xu, Xiaoyu and Jiang, Wenrui and Xu, Lei and Ge, Han and Zhao, Nan and Li, Tiantian and Candini, Andrea and Xi, Bin and Zhao, Jize and Fu, Ying and Yang, Jiong and Zhang, Yuanzhu and Biasiol, Giorgio and Wang, Shanmin and Zhu, Jinlong and Miao, Ping and Tong, Xin and Yu, Dapeng and Mole, Richard and Cui, Yi and Ma, Long and Zhang, Zhitao and Ouyang, Zhongwen and Tong, Wei and Podlesnyak, Andrey and Wang, Ling and Ye, Feng and Yu, Dehong and Yu, Weiqiang and Wu, Liusuo and Wang, Zhentao},
  title   = {{Bose--Einstein condensation of a two-magnon bound state in a spin-1 triangular lattice}},
  journal = {Nature Materials},
  year    = {2025},
  volume  = {24},
  number  = {4},
  pages   = {544--551},
  doi     = {10.1038/s41563-024-02071-z},
  url     = {https://doi.org/10.1038/s41563-024-02071-z},
  issn    = {1476-4660}
}

@article{Xie2023SpinFlip,
  author  = {Xie, Tao and Eberharter, A. A. and Xing, Jie and Nishimoto, S. and Brando, M. and Khanenko, P. and Sichelschmidt, J. and Turrini, A. A. and Mazzone, D. G. and Naumov, P. G. and Sanjeewa, L. D. and Harrison, N. and Sefat, Athena S. and Normand, B. and Läuchli, A. M. and Podlesnyak, A. and Nikitin, S. E.},
  title   = {Complete field-induced spectral response of the spin-1/2 triangular-lattice antiferromagnet CsYbSe2},
  journal = {npj Quantum Materials},
  year    = {2023},
  volume  = {8},
  number  = {1},
  pages   = {48},
  doi     = {10.1038/s41535-023-00580-9},
  url     = {https://doi.org/10.1038/s41535-023-00580-9},
  issn    = {2397-4648}
}

@article{Toth2016Noncollinear1,
  author  = {T{\'o}th, S{\'a}ndor and Wehinger, Bj{\"o}rn and Rolfs, Katharina and Birol, Turan and Stuhr, Uwe and Takatsu, Hiroshi and Kimura, Kenta and Kimura, Tsuyoshi and R{\o}nnow, Henrik M. and R{\"u}egg, Christian},
  title   = {Electromagnon dispersion probed by inelastic X-ray scattering in {LiCrO2}},
  journal = {Nature Communications},
  year    = {2016},
  volume  = {7},
  number  = {1},
  pages   = {13547},
  doi     = {10.1038/ncomms13547},
  url     = {https://doi.org/10.1038/ncomms13547},
  issn    = {2041-1723}
}

@article{Park2016Noncollinear2,
  title = {Magnon-phonon coupling and two-magnon continuum in the two-dimensional triangular antiferromagnet ${\mathrm{CuCrO}}_{2}$},
  author = {Park, Kisoo and Oh, Joosung and Leiner, Jonathan C. and Jeong, Jaehong and Rule, Kirrily C. and Le, Manh Duc and Park, Je-Geun},
  journal = {Phys. Rev. B},
  volume = {94},
  issue = {10},
  pages = {104421},
  numpages = {6},
  year = {2016},
  month = {Sep},
  publisher = {American Physical Society},
  doi = {10.1103/PhysRevB.94.104421},
  url = {https://link.aps.org/doi/10.1103/PhysRevB.94.104421}
}

@article{Rafael2023DynPara,
  title = {Dynamic paramagnon-polarons in altermagnets},
  author = {Steward, Charles R. W. and Fernandes, Rafael M. and Schmalian, J\"org},
  journal = {Phys. Rev. B},
  volume = {108},
  issue = {14},
  pages = {144418},
  numpages = {14},
  year = {2023},
  month = {Oct},
  publisher = {American Physical Society},
  doi = {10.1103/PhysRevB.108.144418},
  url = {https://link.aps.org/doi/10.1103/PhysRevB.108.144418}
}

@article{Weissenhofer2025ChiralPhonons,
  title = {Chiral Phonons Arising from Chirality-Selective Magnon-Phonon Coupling},
  author = {Wei\ss{}enhofer, Markus and Rieger, Philipp and Mrudul, M. S. and Mikadze, Luca and Nowak, Ulrich and Oppeneer, Peter M.},
  journal = {Phys. Rev. Lett.},
  volume = {135},
  issue = {21},
  pages = {216701},
  numpages = {8},
  year = {2025},
  month = {Nov},
  publisher = {American Physical Society},
  doi = {10.1103/j7bs-2zbx},
  url = {https://link.aps.org/doi/10.1103/j7bs-2zbx}
}

@article{Weissenhofer2023RotInv,
  title = {Rotationally invariant formulation of spin-lattice coupling in multiscale modeling},
  author = {Wei\ss{}enhofer, Markus and Lange, Hannah and Kamra, Akashdeep and Mankovsky, Sergiy and Polesya, Svitlana and Ebert, Hubert and Nowak, Ulrich},
  journal = {Phys. Rev. B},
  volume = {108},
  issue = {6},
  pages = {L060404},
  numpages = {6},
  year = {2023},
  month = {Aug},
  publisher = {American Physical Society},
  doi = {10.1103/PhysRevB.108.L060404},
  url = {https://link.aps.org/doi/10.1103/PhysRevB.108.L060404}
}

@article{Kikkawa2016SSE,
  title = {Magnon Polarons in the Spin Seebeck Effect},
  author = {Kikkawa, Takashi and Shen, Ka and Flebus, Benedetta and Duine, Rembert A. and Uchida, Ken-ichi and Qiu, Zhiyong and Bauer, Gerrit E. W. and Saitoh, Eiji},
  journal = {Phys. Rev. Lett.},
  volume = {117},
  issue = {20},
  pages = {207203},
  numpages = {6},
  year = {2016},
  month = {Nov},
  publisher = {American Physical Society},
  doi = {10.1103/PhysRevLett.117.207203},
  url = {https://link.aps.org/doi/10.1103/PhysRevLett.117.207203}
}

@article{Flebus2014MagnonPolaron,
  title = {Magnon-polaron transport in magnetic insulators},
  author = {Flebus, Benedetta and Shen, Ka and Kikkawa, Takashi and Uchida, Ken-ichi and Qiu, Zhiyong and Saitoh, Eiji and Duine, Rembert A. and Bauer, Gerrit E. W.},
  journal = {Phys. Rev. B},
  volume = {95},
  issue = {14},
  pages = {144420},
  numpages = {11},
  year = {2017},
  month = {Apr},
  publisher = {American Physical Society},
  doi = {10.1103/PhysRevB.95.144420},
  url = {https://link.aps.org/doi/10.1103/PhysRevB.95.144420}
}

@article{Chernyshev2006NcAFM,
  title = {Magnon Decay in Noncollinear Quantum Antiferromagnets},
  author = {Chernyshev, A. L. and Zhitomirsky, M. E.},
  journal = {Phys. Rev. Lett.},
  volume = {97},
  issue = {20},
  pages = {207202},
  numpages = {4},
  year = {2006},
  month = {Nov},
  publisher = {American Physical Society},
  doi = {10.1103/PhysRevLett.97.207202},
  url = {https://link.aps.org/doi/10.1103/PhysRevLett.97.207202}
}

@article{Zhang2014EdH,
  title = {Angular Momentum of Phonons and the Einstein--de Haas Effect},
  author = {Zhang, Lifa and Niu, Qian},
  journal = {Phys. Rev. Lett.},
  volume = {112},
  issue = {8},
  pages = {085503},
  numpages = {5},
  year = {2014},
  month = {Feb},
  publisher = {American Physical Society},
  doi = {10.1103/PhysRevLett.112.085503},
  url = {https://link.aps.org/doi/10.1103/PhysRevLett.112.085503}
}

@article{Luo2024PRB,
  title = {Low-temperature spin Seebeck effect in nonmagnetic vanadium dioxide},
  author = {Luo, Renjie and Legvold, Tanner J. and Chen, Liyang and Navarro, Henry and Basaran, Ali C. and Hong, Deshun and Liu, Changjiang and Bhattacharya, Anand and Schuller, Ivan K. and Natelson, Douglas},
  journal = {Phys. Rev. B},
  volume = {110},
  issue = {2},
  pages = {024415},
  numpages = {14},
  year = {2024},
  month = {Jul},
  publisher = {American Physical Society},
  doi = {10.1103/PhysRevB.110.024415},
  url = {https://link.aps.org/doi/10.1103/PhysRevB.110.024415}
}

@article{Luo2024PRBnoise,
  title = {{Challenges of measuring spin Seebeck noise}},
  author = {Luo, Renjie and Zhao, Xuanhan and Legvold, Tanner J. and Chen, Liyang and Liu, Changjiang and Hong, Deshun and Bhattacharya, Anand and Natelson, Douglas},
  journal = {Phys. Rev. B},
  volume = {109},
  issue = {10},
  pages = {104429},
  numpages = {9},
  year = {2024},
  month = {Mar},
  publisher = {American Physical Society},
  doi = {10.1103/PhysRevB.109.104429},
  url = {https://link.aps.org/doi/10.1103/PhysRevB.109.104429}
}

@article{Luo2023APL,
    author = {Luo, Renjie and Legvold, Tanner J. and Chen, Liyang and Natelson, Douglas},
    title = {{Nernst–Ettingshausen effect in thin Pt and W films at low temperatures}},
    journal = {Applied Physics Letters},
    volume = {122},
    number = {18},
    pages = {182405},
    year = {2023},
    month = {05},
    issn = {0003-6951},
    doi = {10.1063/5.0146427},
    url = {https://doi.org/10.1063/5.0146427}
}

@article{Luo2025PRB,
  title = {{Spin Seebeck effect in correlated antiferromagnetic ${\mathrm{V}}_{2}{\mathrm{O}}_{3}$}},
  author = {Luo, Renjie and Legvold, Tanner J. and Eichman, Gage and Navarro, Henry and Basaran, Ali C. and Qiu, Erbin and Schuller, Ivan K. and Natelson, Douglas},
  journal = {Phys. Rev. B},
  volume = {112},
  issue = {18},
  pages = {184421},
  numpages = {8},
  year = {2025},
  month = {Nov},
  publisher = {American Physical Society},
  doi = {10.1103/7xhk-3pc2},
  url = {https://link.aps.org/doi/10.1103/7xhk-3pc2}
}

@article{Kim2023chiralPhonon,
  author  = {Kim, Kyunghoon and Vetter, Eric and Yan, Liang and Yang, Cong and Wang, Ziqi and Sun, Rui and Yang, Yu and Comstock, Andrew H. and Li, Xiao and Zhou, Jun and Zhang, Lifa and You, Wei and Sun, Dali and Liu, Jun},
  title   = {Chiral-phonon-activated spin Seebeck effect},
  journal = {Nature Materials},
  year    = {2023},
  volume  = {22},
  number  = {3},
  pages   = {322--328},
  doi     = {10.1038/s41563-023-01473-9},
  url     = {https://doi.org/10.1038/s41563-023-01473-9},
  issn    = {1476-4660}
}

@article{Juraschek2025ChiralPhonons,
  author  = {Juraschek, Dominik M. and Geilhufe, R. Matthias and Zhu, Hanyu and Basini, Martina and Baum, Peter and Baydin, Andrey and Chaudhary, Swati and Fechner, Michael and Flebus, Benedetta and Grissonnanche, Gael and Kirilyuk, Andrei I. and Lemeshko, Mikhail and Maehrlein, Sebastian F. and Mignolet, Maxime and Murakami, Shuichi and Niu, Qian and Nowak, Ulrich and Romao, Carl P. and Rostami, Habib and Satoh, Takuya and Spaldin, Nicola A. and Ueda, Hiroki and Zhang, Lifa},
  title   = {Chiral phonons},
  journal = {Nature Physics},
  year    = {2025},
  volume  = {21},
  number  = {10},
  pages   = {1532--1540},
  doi     = {10.1038/s41567-025-03001-9},
  url     = {https://doi.org/10.1038/s41567-025-03001-9},
  issn    = {1745-2481}
}

@article{Barkeshli2012HallVisc,
  title = {Dissipationless phonon Hall viscosity},
  author = {Barkeshli, Maissam and Chung, Suk Bum and Qi, Xiao-Liang},
  journal = {Phys. Rev. B},
  volume = {85},
  issue = {24},
  pages = {245107},
  numpages = {12},
  year = {2012},
  month = {Jun},
  publisher = {American Physical Society},
  doi = {10.1103/PhysRevB.85.245107},
  url = {https://link.aps.org/doi/10.1103/PhysRevB.85.245107}
}

@article{Flebus2023HallVisc,
  title = {Phonon Hall Viscosity of Ionic Crystals},
  author = {Flebus, B. and MacDonald, A. H.},
  journal = {Phys. Rev. Lett.},
  volume = {131},
  issue = {23},
  pages = {236301},
  numpages = {6},
  year = {2023},
  month = {Dec},
  publisher = {American Physical Society},
  doi = {10.1103/PhysRevLett.131.236301},
  url = {https://link.aps.org/doi/10.1103/PhysRevLett.131.236301}
}

@article{Avron1995HallVisc,
  title = {Viscosity of Quantum Hall Fluids},
  author = {Avron, J. E. and Seiler, R. and Zograf, P. G.},
  journal = {Phys. Rev. Lett.},
  volume = {75},
  issue = {4},
  pages = {697--700},
  numpages = {0},
  year = {1995},
  month = {Jul},
  publisher = {American Physical Society},
  doi = {10.1103/PhysRevLett.75.697},
  url = {https://link.aps.org/doi/10.1103/PhysRevLett.75.697}
}

@article{Ren2024Adiabatic,
  title = {Adiabatic Dynamics of Coupled Spins and Phonons in Magnetic Insulators},
  author = {Ren, Shang and Bonini, John and Stengel, Massimiliano and Dreyer, Cyrus E. and Vanderbilt, David},
  journal = {Phys. Rev. X},
  volume = {14},
  issue = {1},
  pages = {011041},
  numpages = {30},
  year = {2024},
  month = {Mar},
  publisher = {American Physical Society},
  doi = {10.1103/PhysRevX.14.011041},
  url = {https://link.aps.org/doi/10.1103/PhysRevX.14.011041}
}

@article{Bethe1931TwoMag,
  author  = {Bethe, H.},
  title   = {Zur Theorie der Metalle},
  journal = {Zeitschrift f{\"u}r Physik},
  year    = {1931},
  volume  = {71},
  number  = {3},
  pages   = {205--226},
  doi     = {10.1007/BF01341708},
  url     = {https://doi.org/10.1007/BF01341708}
}

@article{Wortis1963TwoMag,
  title = {Bound States of Two Spin Waves in the Heisenberg Ferromagnet},
  author = {Wortis, Michael},
  journal = {Phys. Rev.},
  volume = {132},
  issue = {1},
  pages = {85--97},
  numpages = {0},
  year = {1963},
  month = {Oct},
  publisher = {American Physical Society},
  doi = {10.1103/PhysRev.132.85},
  url = {https://link.aps.org/doi/10.1103/PhysRev.132.85}
}

@article{Torrance1969LocalTwoMag,
  title = {Magnon Bound States in Anisotropic Linear Chains},
  author = {Torrance, J. B. and Tinkham, M.},
  journal = {Phys. Rev.},
  volume = {187},
  issue = {2},
  pages = {587--594},
  numpages = {0},
  year = {1969},
  month = {Nov},
  publisher = {American Physical Society},
  doi = {10.1103/PhysRev.187.587},
  url = {https://link.aps.org/doi/10.1103/PhysRev.187.587}
}

@article{Fukuhara2013LocalFlip,
  author  = {Fukuhara, Takeshi and Schau{\ss}, Peter and Endres, Manuel and Hild, Sebastian and Cheneau, Marc and Bloch, Immanuel and Gross, Christian},
  title   = {Microscopic observation of magnon bound states and their dynamics},
  journal = {Nature},
  year    = {2013},
  volume  = {502},
  number  = {7469},
  pages   = {76--79},
  doi     = {10.1038/nature12541},
  url     = {https://doi.org/10.1038/nature12541},
  issn    = {1476-4687}
}

@misc{Williams2026ChemPot,
      title={Chemical potential of magnon polarons}, 
      author={Violet Williams and Benedetta Flebus},
      year={2026},
      archivePrefix={arXiv},
      primaryClass={cond-mat.mes-hall},
      url={https://arxiv.org/abs/2512.02222}, 
}

@article{Coh2023PT,
  title = {Classification of materials with phonon angular momentum and microscopic origin of angular momentum},
  author = {Coh, Sinisa},
  journal = {Phys. Rev. B},
  volume = {108},
  issue = {13},
  pages = {134307},
  numpages = {8},
  year = {2023},
  month = {Oct},
  publisher = {American Physical Society},
  doi = {10.1103/PhysRevB.108.134307},
  url = {https://link.aps.org/doi/10.1103/PhysRevB.108.134307}
}

@article{Coh2023ChiralPhonon,
  title = {{Frequency Splitting of Chiral Phonons from Broken Time-Reversal Symmetry in ${\mathrm{CrI}}_{3}$}},
  author = {Bonini, John and Ren, Shang and Vanderbilt, David and Stengel, Massimiliano and Dreyer, Cyrus E. and Coh, Sinisa},
  journal = {Phys. Rev. Lett.},
  volume = {130},
  issue = {8},
  pages = {086701},
  numpages = {6},
  year = {2023},
  month = {Feb},
  publisher = {American Physical Society},
  doi = {10.1103/PhysRevLett.130.086701},
  url = {https://link.aps.org/doi/10.1103/PhysRevLett.130.086701}
}

@article{ralko2026orderbydisorder,
  title = {Order-by-disorder from Schwinger bosons in a frustrated honeycomb ferromagnet},
  author = {Ralko, Arnaud and Merino, Jaime},
  journal = {Phys. Rev. B},
  volume = {113},
  issue = {21},
  pages = {214413},
  numpages = {13},
  year = {2026},
  month = {Jun},
  publisher = {American Physical Society},
  doi = {10.1103/7h87-lgv2},
  url = {https://link.aps.org/doi/10.1103/7h87-lgv2}
}

@article{gao2025neutron,
  title={Neutron scattering and thermodynamic evidence for emergent photons and fractionalization in a pyrochlore spin ice},
  author={Gao, Bin and Desrochers, F{\'e}lix and Tam, David W and Kirschbaum, Diana M and Steffens, Paul and Hiess, Arno and Nguyen, Duy Ha and Su, Yixi and Cheong, Sang-Wook and Paschen, Silke and Kim, Yong-Baek and Dai, Pengcheng},
  journal={Nature Physics},
  volume={21},
  number={8},
  pages={1203--1210},
  year={2025},
  publisher={Nature Publishing Group UK London}
}

\end{document}